\documentclass[prlprintnumbers,showpacs,reprint,prl,twocolumn,floatfix,superscriptaddress,amsmath,amssymb]{revtex4-1}
\usepackage{epsfig}
\usepackage{subfigure}
\usepackage{amsmath}
\usepackage[usenames,dvipsnames]{color}
\usepackage{amssymb}
\usepackage{setspace}
\usepackage{graphicx} 
\usepackage{dcolumn}
\usepackage{bm}
\usepackage{times}
\input epsf
\usepackage{hyperref}
\hypersetup{
    bookmarks=true,         
    unicode=false,          
    pdftoolbar=true,        
    pdfmenubar=true,        
    pdffitwindow=false,     
    pdfstartview={FitH},    
    pdftitle={My title},    
    pdfauthor={Daniel Larremore},     
    pdfsubject={Research Journal},   
    pdfcreator={},   
    pdfproducer={}, 
    pdfkeywords={} {} {}, 
    pdfnewwindow=true,      
    colorlinks=true,       
    linkcolor=red,          
    citecolor=Green,        
    filecolor=magenta,      
    urlcolor=blue           
}

\begin{document}
\author{Daniel B. Larremore}
\affiliation{Department of Epidemiology, Harvard School of Public Health, Boston, MA, 02115, USA}
\affiliation{Center for Communicable Disease Dynamics, Harvard School of Public Health, Boston, MA, 02115, USA}

\author{Aaron Clauset}
\affiliation{Department of Computer Science, University of Colorado, Boulder, CO, 80303, USA}
\affiliation{BioFrontiers Institute, University of Colorado, Boulder, CO, 80303, USA}
\affiliation{Santa Fe Institute, Santa Fe, NM, 87501, USA}

\author{Caroline Buckee}
\affiliation{Department of Epidemiology, Harvard School of Public Health, Boston, MA, 02115, USA}
\affiliation{Center for Communicable Disease Dynamics, Harvard School of Public Health, Boston, MA, 02115, USA}

\begin{abstract}
The {\it var} genes of the human malaria parasite {\it Plasmodium falciparum} present a challenge to population geneticists due to their extreme diversity, which is generated by high rates of recombination. These genes encode a primary antigen protein called PfEMP1, which is expressed on the surface of infected red blood cells and elicits protective immune responses. {\it var} gene sequences are characterized by pronounced mosaicism, precluding the use of traditional phylogenetic tools that require bifurcating tree-like evolutionary relationships. We present a new method that identifies highly variable regions (HVRs), and then maps each HVR to a complex network in which each sequence is a node and two nodes are linked if they share an exact match of significant length. Here, networks of {\it var} genes that recombine freely are expected to have a uniformly random structure, but constraints on recombination will produce network communities that we identify using a stochastic block model. We validate this method on synthetic data, showing that it correctly recovers populations of constrained recombination, before applying it to the Duffy Binding Like-$\alpha$ (DBL$\alpha$) domain of {\it var} genes. We find nine HVRs whose network communities map in distinctive ways to known DBL$\alpha$ classifications and clinical phenotypes. We show that the recombinational constraints of some HVRs are correlated, while others are independent. These findings suggest that this micromodular structuring facilitates independent evolutionary trajectories of neighboring mosaic regions, allowing the parasite to retain protein function while generating enormous sequence diversity. Our approach therefore offers a rigorous method for analyzing evolutionary constraints in {\it var} genes, and is also flexible enough to be easily applied more generally to any highly recombinant sequences. 
\end{abstract}

\title{A network approach to analyzing highly recombinant malaria parasite genes}

\maketitle

\section{Author Summary}
The human malaria parasite kills nearly 1 million people each year globally. Frequent genetic exchange between malaria parasites creates enormous genetic diversity that largely explains the lack of an effective vaccine for the disease. Traditional phylogenetic tools cannot accommodate this type of diversity, however, and rigorous analytical tools capable of making sense of gene sequences that recombine frequently are still lacking.  Here, we use network techniques that have been developed by the physics and network science communities to analyze malaria parasite gene sequences, allowing us to automatically identify highly variable mosaic regions in sequence data and to derive the network of recombination events. We apply our method to seven fully-sequenced parasite genomes, and show that our method provides new insights into the complex evolutionary patterns of the parasite. Our results suggest that the structure of these sequences allows the parasite to rapidly diversify to evade immune responses while maintaining antigen structure and function.

\section{Introduction}
The human malaria parasite {\it Plasmodium falciparum} causes approximately 1 million deaths each year, primarily in young children in sub-Saharan Africa \cite{1}. In endemic regions, individuals develop clinical immunity to severe disease in childhood, but continue to suffer malaria infections and mild illness throughout their lifetimes. This epidemiological pattern is poorly understood, but appears to be caused by the gradual acquisition of a large repertoire of antibodies following sequential exposure to different parasite proteins \cite{2,3,4,5}. The main candidate for eliciting protective antibodies is the parasite-derived antigen PfEMP1 ({\it P. falciparum} erythrocyte membrane protein 1), encoded in each parasite genome by a large {\it var} gene family and expressed during infection on the surface of infected red blood cells in a process of antigenic variation \cite{2,6,7,8,9}. Extremely rapid recombination among {\it var} genes generates enormous diversity and complex mosaic structures among these sequences \cite{10,11,12,13,14}, and recent field studies have uncovered seemingly limitless {\it var} gene diversity in Africa \cite{15}. Superinfection with multiple clones is extremely common, and recombination can occur during meiosis in the mosquito, as well as between {\it var} genes on different chromosomes of a single parasite during asexual reproduction \cite{14}. However, these observations are at odds with the rapid acquisition of antibodies to common PfEMP1 variants that are associated with disease \cite{16}, as well as the finding that parasites from different continents share identical sequence blocks despite millions of years of evolutionary separation \cite{17}.

The highly recombinant structure of {\it var} genes precludes the use of standard phylogenetic tools, and the processes generating this paradoxical relationship between parasite genetic structure and the epidemiology of infection and disease remain unclear \cite{13,18,19}. Statistically rigorous and scalable techniques to analyze evolutionary relationships between sequences generated through frequent recombination are lacking. Classical phylogenetic analyses are designed to accommodate branching tree-like relationships between genes generated by mutation, and therefore require that highly recombinant regions, where evolutionarily distant sequences may share mosaics, are removed, ignored, or assumed to be absent \cite{20,21,22}. Bockhorst et al. introduced an approach to understanding the most conserved group of {\it var} genes based on a segmentation analysis which divides a set of sequences into segments such that polymorphic sites in the same segment are strongly correlated, while nearby polymorphic sites are either weakly or not correlated \cite{18,23}. While segmentation analysis is useful to detect mutation-driven diversification following ancient recombination or geographic separation, particularly for subsets of more conserved {\it var} genes, it ultimately generates a tree-like relationship between genes and does not accommodate recent and ongoing recombination.     

Networks provide a mathematical approach to representing and studying complex relationships between genes \cite{24}, and network-based techniques have produced valuable evolutionary insights for many organisms ranging from viruses to eukaryotes \cite{25}. Attempts to introduce recombination within phylogenetic frameworks have led to specialized techniques that produce phylogenetic (or recombination) networks for small numbers of sequences when recombination rates are relatively low, but these also focus on conserved regions rather than providing insights into the recombinant regions \cite{26} (for a review see \cite{27}). On the other hand, ancestral recombination graphs have a strong theoretical foundation but lack efficient approximations that are required for rapid inference \cite{28}. Networks have also been used to identify large-scale clusters of global gene sharing and exchange \cite{29} and horizontal gene transfer of the plasmid resistome \cite{30}, as well as differentiating horizontal and vertical flow of information in \cite{31,32}. In these approaches, all-to-all BLAST scores are calculated and thresholded for a set of sequences, and the resulting network is generally analyzed visually to assess large-scale structure \cite{22,25,29,30,31,32}. These analyses, however, rely on ad hoc parameter choices, uncontrolled assumptions, or prior knowledge of target clustering. And, while potentially useful for hypothesis generation, a reliance on alignment scores contains an implicit model for sequence mutation and substitution that is not justified for highly recombinant {\it var} gene sequences. We have previously taken a network approach to analyze clinical {\it var} gene domains using short position-specific sequences \cite{13}. Although this approach uncovered distinctive structuring, with clustering that reflected previous {\it var} gene classification schemes, it lacked a solid theoretical basis and more importantly it was not generalizable to other domains and genes.

Here we take a more sophisticated approach, applying rigorous community detection methods that have primarily been developed in the physics, statistics, and network science literature, to construct and analyze recombinant gene networks in general, and {\it var} gene networks in particular. We apply our technique to previously published and annotated sequences of the {\it var} Duffy Binding Like-$\alpha$ (DBL$\alpha$) domain \cite{18,33}, which unlike other domains is found in almost all {\it var} genes sequenced. We show that networks constructed from different mosaic regions across the domain vary widely in their community structure, uncovering a new layer of micromodularity among {\it var} genes. Our results imply a lack of coupled evolution within even a single domain. At the same time, clear structures within networks correspond well, and differentially, to previously published classifications that have been linked to disease phenotypes. This structuring therefore provides a mechanism to generate vast diversity while maintaining protein structure and function, reconciling the paradoxical observations of both common serological responses and almost limitless {\it var} sequence diversity.  

\section{Methods}
\subsection{{\it Plasmodium falciparum} {\it var}  gene data }
We analyze 307 amino acid sequences from the DBL$\alpha$ domain of the {\it var} genes of seven {\it P. falciparum} isolates published in \cite{18}. PfEMP1 antigens exhibit modular structures, characterized by between two and nine DBL and CIDR (Cysteine-Rich Interdomain Region) domains \cite{18,34,35,36}. While there are many different classes of these domains, indexed by $\alpha$, $\beta$, etc., the N-terminal region of the protein almost always begins with a DBL$\alpha$ and CIDR$\alpha$ pair, each of which has been implicated in the binding of infected red blood cells to various host receptors as well as different disease pathologies. To highlight the diversity and mosaicism of the DBL$\alpha$ domain we first apply standard phylogenetic approaches to the 307 sequences. Sequence length prior to alignment was widely distributed between 357 and 473 amino acids with median 420 and mode 398. Pair-wise alignments using the standard tool MUSCLE \cite{16,37} averaged 5.6\% (or 23) gaps. Due to the presence of highly variable regions, a multiple alignment required an implausibly large number of gap insertions, yielding an aligned length of 743. The remarkable diversity in DBL$\alpha$ sequences is further illustrated in the unresolved nature of a phylogenetic tree built from such an alignment (Figure \ref{figs3}).

\begin{figure*}
	\includegraphics[width=0.95\linewidth]{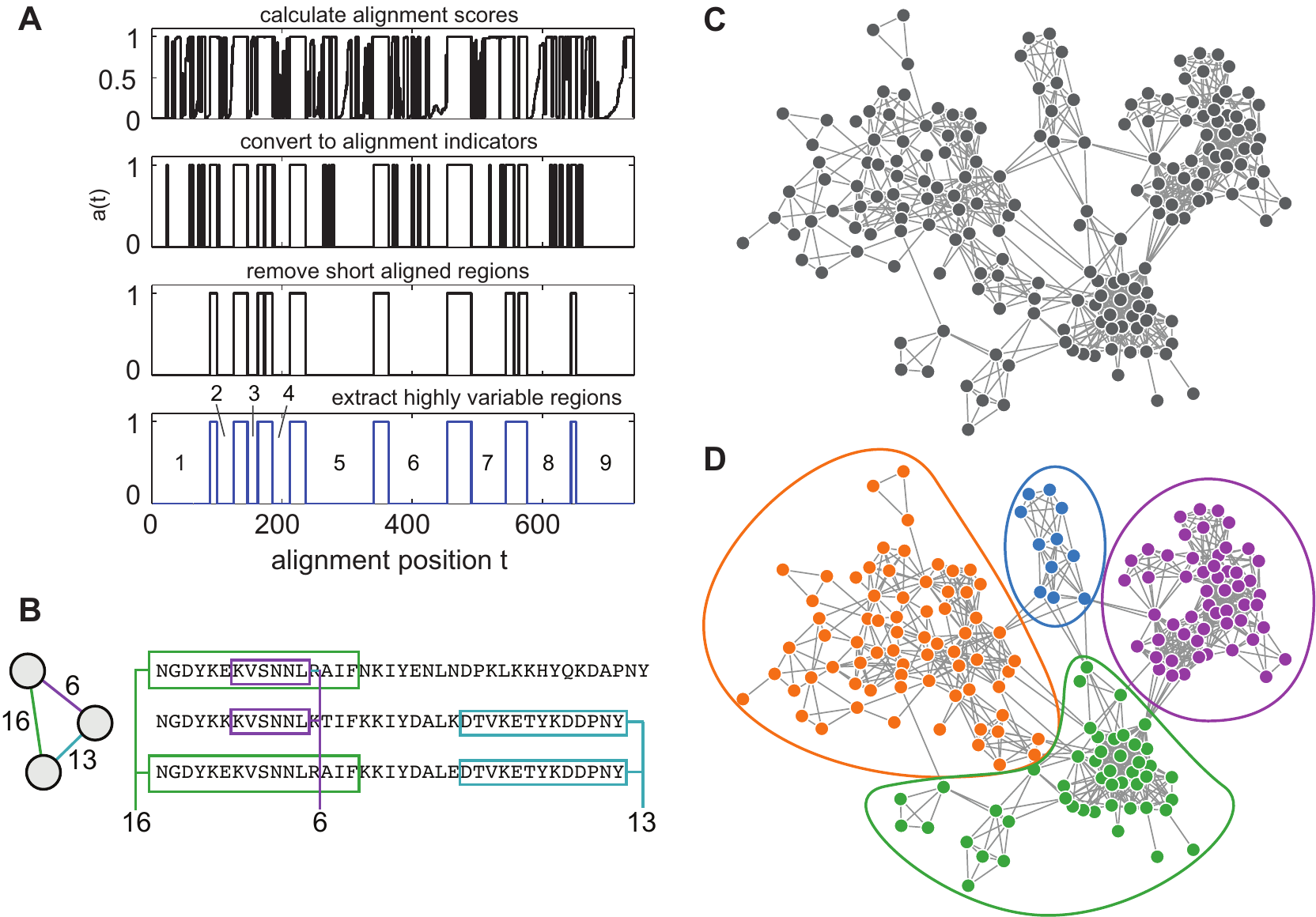}
	\caption{{\bf Pictorial overview of sequence analysis method.} (A) Starting from a multiple alignment of the domain of interest, four steps are taken to identify highly variable regions (HVRs), as described in the text. We show the HVR identification process for the 307 DBL$\alpha$ sequences from \cite{18}.  (B) For each HVR, a network is made in which each sequence is a node, and a link connecting two nodes corresponds to a shared sequence block. (C) The set of pairwise connections above the noise threshold defines a complex network representing recent recombination events. (D) Communities are inferred directly from this network using a probabilistic generative model. Steps B,C, and D are repeated for each of the HVRs identified in step A.}
	\label{fig1}
\end{figure*}

\subsection{General approach}
Instead of using alignments to identify evolutionary signals contained in highly conserved regions, we use them to identify and remove conserved regions in order to focus on recombinant mosaic sequences. The method uses three steps, which we motivate here, and define in detail in the next sections: (i) Identify highly variable regions (HVRs) across all sequences. (ii) Compare sequences pair-wise within each HVR, generating a distinct block-sharing network for each region. (iii) Statistically identify communities in each network, which will represent groups of {\it var} genes that recombine more frequently with each other than with genes from other communities. These general steps are illustrated in Figure \ref{fig1}.

\begin{figure*}
	\includegraphics[width=0.95\linewidth]{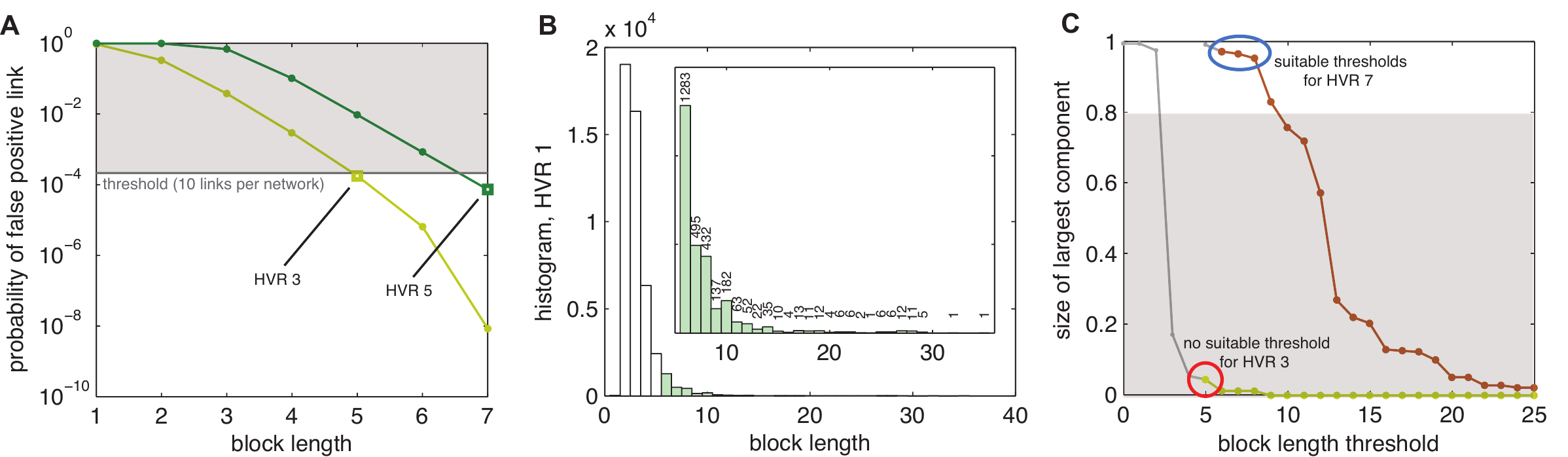}
	\caption{{\bf Choice of link noise threshold.} Choosing a noise threshold requires balance between two competing requirements for correctly identifying network communities: minimize the number of incorrectly placed links, yet retain as many correctly placed links as possible to satisfy the network connectivity requirements of the community detection method. (A) The probability of two sequences sharing a block while not actually being related decreases as block length increases, modeled in S1. Each HVRÕs length and composition are taken into account separately (colored lines). Choosing a tolerance for false positives (grey line) specifies a minimum retained block length; since blocks are of integer length, the next largest integer is the minimum retained block length (squares). Curves for HVRs 3 and 5 are plotted, for which we would select thresholds of five and seven, respectively. Curves for all nine HVRs are shown in Fig. S6A. (B) For a choice of threshold 6 for HVR 1, the histogram of HVR 1 block lengths shows that a vast majority of the blocks are below the threshold (white bars) and that the retained blocks are widely distributed (green bars, inset). (C) Networks are fragmented as the block length threshold is increased and more links are discarded. The relationship between the size of the largest component and block length threshold is shown for the least-connected (HVR3) and most-connected (HVR7) networks. Some thresholds allow too many false positives, as described in panel A (grey lines), yet other thresholds fragment the network too much for reliable community detection (shaded region). Those points that are plotted in color above the shaded region are both sufficiently error-free and well connected that we may reliably infer network communities. For HVRs 2-4, even the most permissive false positive threshold results in a network that is too fragmented for community detection (red circle). Curves for all nine HVRs are shown in Figure \ref{figs6}B.}
	\label{fig2}
\end{figure*}

To identify HVRs, we use the basic premise of an alignment as a starting point: highly variable regions will require gap insertions in order to find an alignment. In contrast with other methods used to identify regions of conservation by discarding poorly-aligned stretches \cite{38}, our explicit goal is to find contiguous poorly-aligned regions, since these are likely to be mosaics resulting from recombination. After identifying HVRs, instead of constructing trees, we generate a complex network for each HVR using an alignment-free process, where each vertex is a sequences and two sequences are connected if they exhibit a pattern of recombination. The structure of this network reflects the constraints and extents of the recombination process. Since any two genes may recombine in the absence of constraints on recombination, deviations from a random network represent structured recombination, function, or evolution between {\it var} gene communities. To analyze network structures, we use a community-detection approach that can identify the patterns produced by constrained recombination, by fitting a generative model called a degree-corrected stochastic block model \cite{39} to the network data. The degree-corrected stochastic block model identifies communities by picking out non-random patterns in the network connections making it an appropriate choice among myriad community detection methods. Each step is described in detail below. 

\subsection{Detailed approach}
\subsubsection{Defining Highly Variable Regions (HVRs)}
In this step, we take a set of amino acid sequences and identify highly variable regions. Starting from a multiple alignment, each aligned position t is first assigned an alignment score representing the fraction of input sequences that are aligned at that position (i.e. not gap insertions). This score is used to calculate an alignment indicator a(t) such that when all sequences align with no gaps at position t, a(t)=1; if there are any gaps, a(t)=0. We then identify regions where the sequences align for G or more consecutive positions, that is, a(t)=1 for G or more consecutive t. These well aligned regions will serve as separators between HVRs. We choose G sufficiently short, based on a simple null model of sequences (SI1), that any highly conserved block of significant length is removed from network construction since it would obscure patterns of recombination. Thus, the HVRs will be the regions in between the conserved regions that we have just identified. However, very short HVRs will have so few amino acids once gaps are removed in subsequent steps that they are unlikely to reflect the mosaicism in which we are interested. So, we define a minimum HVR size H, discarding those that are shorter. These steps are illustrated in Figure \ref{fig1}A. 

\subsubsection{Generating a recombination network}
Next, we take each HVR and produce an unweighted and undirected network. Each node represents a sequence, and each link represents a shared sequence block, indicating a recombinant relationship between the two sequences. Before comparing sequences, all gap insertions from the alignment process are removed. First, we create a weighted network in which the weight of each link is the length of the longest substring shared between the sequences it connects (Figure \ref{fig1}B). The result is an all-to-all undirected and weighted network for each HVR. Thus, for sequences with multiple HVRs there will be multiple networks. Next, we convert each all-to-all weighted network into a sparse and unweighted network by discarding links with weight below a threshold, and removing weights from the remaining links (Figure \ref{fig1}C). Here we choose the threshold in a way that controls the number of false positive links that may have arisen by chance, as shown in Figures \ref{fig2}A and \ref{fig2}B.  The method for computing a noise threshold is based on a null model for randomly assembled sequences using the properties of each HVR, and not derived from network properties. Thus, depending on the confidence one wishes to have in the validity of the networkÕs links, a threshold may be computed from a selected tolerable error rate. Derivation of the function used for this computation is included in supplement S1. 

\subsubsection{Finding recombination communities}
In the final step, we take an unweighted and undirected HVR network and apply a degree-corrected stochastic block model \cite{39} to identify community structures, illustrated by Figure \ref{fig1}D. This model takes as its input an unweighted, undirected network and the number of communities k for which it should find a maximum likelihood fit, and provides as an output a list of which nodes belong to which of the k communities, also referred to as a partition. (Derivation and maximization of the likelihood function are discussed at length in Ref. \cite{39} and efficient code has been made publicly available by Karrer and Newman.) Because previous classifications included between three and six types, we inferred community structures for k=3 to k=6. Each HVR network may have different community structure, similar to how in a standard approach, different loci may generate different phylogenetic trees. However, clades in trees represent distinct branches in an evolutionary history, while HVR network communities represent distinct clusters of ongoing recombination. Multiple trees may be combined to produce a consensus tree, but HVR networks show no clear consensus. We compare our resulting network communities to previous analyses of {\it var} gene sequence groups. Weights are removed from the network prior to community detection for two reasons. First, it is unclear by what principle differences in weights should be interpreted when defining communities. Second, the problem of correctly inferring degree-corrected stochastic block model community structure in sparse and weighted networks is currently unsolved. For these reasons, we interpret each network link as evidence of some recombinant or hereditary history and treat them equally by unweighting networks prior to community detection. We validate this approach carefully as follows. 

\begin{figure}
	\includegraphics[width=0.95\linewidth]{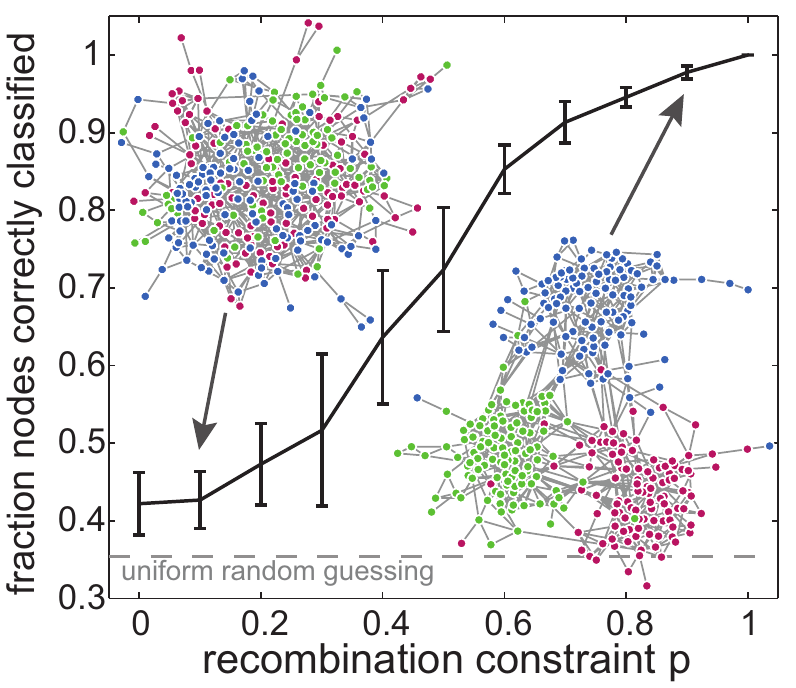}
	\caption{{\bf Performance on synthetic data.} We validate our method's ability to detect constraints on recombination by testing it on synthetic data with known structure. Sequences were generated at random and divided into three communities, after which 1000 recombination events were simulated, described fully in S2. For each recombination event, the two sequences were forced to be chosen from the same community with probability p or were selected uniformly at random with probability $(1-p)$. As the probability that recombination is constrained to within-community is varied from no constraint $(p=0)$ to strict constraint $(p=1)$, the ability of our method to correctly classify sequences into one of three communities increases from very poor to perfect. The connected line shows the mean of 25 replicates, with whiskers indicating $\pm$ one standard deviation. Two example networks are shown for $p=0.1$ and $p=0.9$. The dashed line indicates the accuracy of guessing communities uniformly at random, which is slightly larger than 1/3 as explained in S2. Networks are displayed using a force-directed algorithm that allows a system of repelling point-charges (nodes) and linear springs (links) to relax to a low-energy two dimensional configuration, allowing for visualization of network communities.}
	\label{fig3}
\end{figure}

\subsubsection{Validation on synthetic data}
In order to confirm that our method is able to correctly recover recombinant communities, we create synthetic sequences with varying constraints on recombination between predefined groups. We begin by creating amino acid sequences at random from an empirical amino acid frequency distribution and separating them arbitrarily into three groups. Then, we simulate recombination events in which two parent sequences recombine to produce a child sequence, inheriting the group label of one of its parents. We first choose a parent uniformly at random from the population. Then, with probability $p$, the other parent is chosen from the same group and with probability $(1-p)$ the other parent is chosen uniformly at random. As $p$ is increased from zero to one, the rate of inter-group recombination goes to zero; the communities within the networks produced by applying our sequence analysis method to the synthetically recombined sequences become more well defined, and the method becomes increasingly accurate in correctly classifying nodes. As shown in Figure \ref{fig3}, our method is able to recover recombinant communities perfectly in the presence of strong constraints, and performs only slightly better than random guessing when there are no constraints, as expected. Details of the validation process are found in SI3. 

\begin{figure*}
	\includegraphics[width=0.8\linewidth]{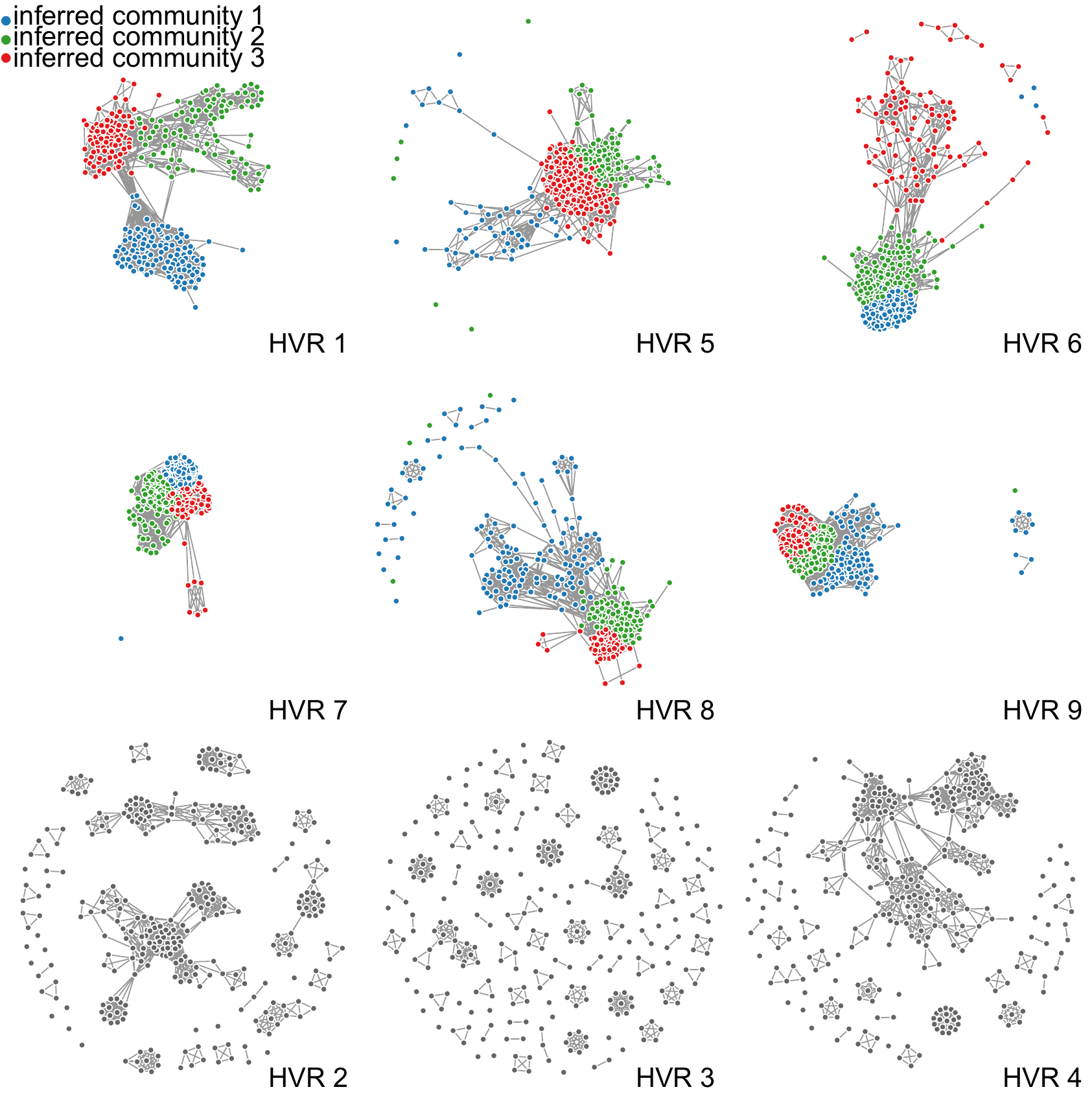}
	\caption{{\bf Nine HVR networks colored by inferred communities.} DBL$\alpha$ HVR networks show a wide range of characteristics, including community size, number of components, and number of links. Nodes in each HVR are colored according to the best three-community partition identified by the inference algorithm (see text). (Identical networks colored by upstream promotor are found in Figure \ref{figs2}.) HVRs 2-4 are sufficiently fragmented that block model inference cannot be trusted. An interactive version of this figure with varying communities and node labels may be found at \href{http://danlarremore.com/var}{http://danlarremore.com/var}. Networks are displayed using a force-directed algorithm that allows a system of repelling point-charges (nodes) and linear springs (links) to relax to a low-energy two dimensional configuration, allowing for visualization of network communities.}
	\label{fig4}
\end{figure*}

\begin{table*}
\caption{Summary statistics for DBL$\alpha$ HVRs.}
\label{table1}
\begin{tabular}{ l | l l l l l l l l l }
HVR	 & 1&	2&	3&	4&	5&	6&	7&	8&	9 \\
\hline \hline
Med. length & 28 & 11 & 6 & 12 & 42 & 29 & 37 & 35 & 39 \\
(min-max) &(22-48) &(8-17) &(4-9) &(6-15) &(30-57) &(25-47) &(16-40) &(21-49) &(8-76) \\
\hline
Length incl. gaps&	88&	27&	17&	28&	106&	91&	54&	68&	92 \\
\hline
Noise cutoff &	6&	6&	5&	5&	7&	6&	6&	7&	6 \\
\hline
\% significant links &	6.0&	3.1&	1.5&	2.6&	5.8&	6.9&	24.9&	8.4&	16.1 \\
\hline
\end{tabular}
\end{table*}

\begin{table*}
\caption{Summary statistics for DBL$\alpha$ HVR networks.}
\label{table2}
\begin{tabular}{ l | l l l l l l l l l }
HVR Network	&1&	2&	3&	4&	5&	6&	7&	8&	9 \\
\hline \hline
N. components &	1&	39&	106&	51&	10&	8&	2&	20&	4\\
\hline
Largest comp. (307)&	307&	112&	18&	183&	298&	291&	306&	273&	293 \\
\hline
\end{tabular}
\end{table*}


\section{Results and Discussion}
\subsection{Discordant patterns of recombination preclude a consensus network}

\subsubsection{Characteristics and community structures of HVR networks}
Nine HVRs were found in the 307 DBL$\alpha$ sequences \cite{18}, using HVR detection parameters of $G=8$ and $H=6$. These parameters were chosen based on the previously described model for false positive links (SI2), and HVR boundaries were not dramatically affected by small changes to these parameters. The nine HVRs found here corresponded partially to previously identified variable regions over all DBL domains \cite{33}. Since HVRs are by definition highly variable, they consist of mostly gap insertionsÑthis diversity is highlighted by the fact that when gaps were removed after HVRs were identified, sequences shrank by 57\% on average. Noise thresholds were computed as shown in Figure \ref{fig2}A such that, in expectation, 10 links (0.02\%) or less are false positives, yielding cutoff lengths of 5, 6, or 7 amino acids, varying by HVR. Each HVR showed a wide range of sequence lengths, and HVRs differed widely from each other in median sequence length. HVR lengths, noise cutoffs, and the percentage of links retained for community detection are found in Table \ref{table1}. The fraction of links above the length cutoffs varied by HVR, shown in Figure \ref{fig2}B and Table \ref{table1}. For HVRs 2-4, removal of links below the cutoff fragmented the network and in such cases community structure cannot be inferred. Figure \ref{fig2}C illustrates graphically that there did not exist a threshold that both preserved a large connected component and met our requirements for a low false positive rate for HVRs 2-4.

\begin{figure*}
	\includegraphics[width=0.95\linewidth]{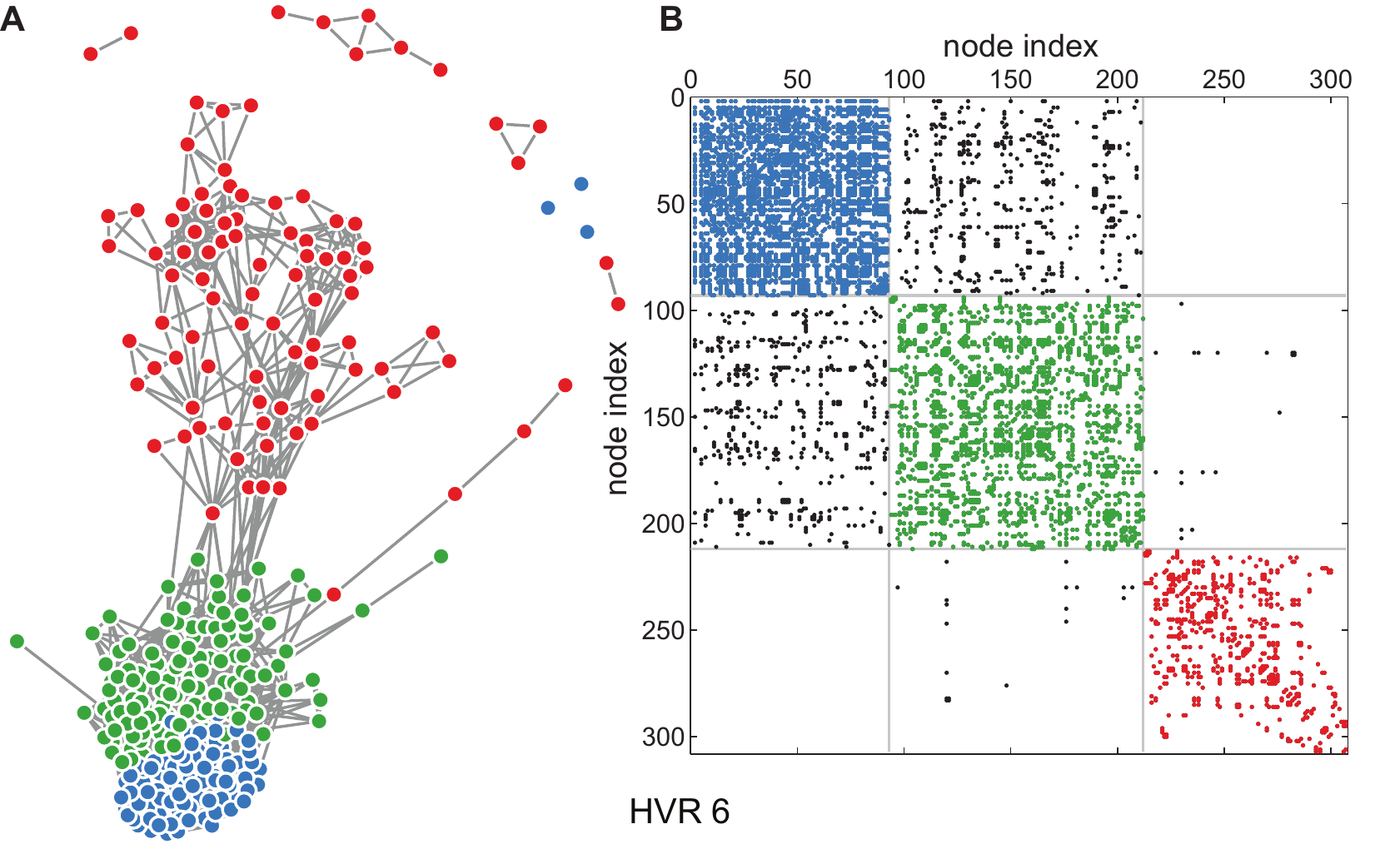}
	\caption{{\bf Stochastic block modeling identifies network comunities.} HVR 6 is shown in two forms, colored according to the best partition into three communities. (A) A force-directed visualization of the network with the identified communities labeled by color. (B) Adjacency matrix in which the ordering of rows and columns has been permuted to match the inferred communities. Diagonal colored blocks are within-community links, and off-diagonal blocks are between-community links. The matrix is shown symmetrically to aid the eye.}
	\label{fig5}
\end{figure*}

Each remaining HVR network had identifiable communities, examples of which are illustrated in Figures \ref{fig4} and \ref{fig5}. This is consistent with our previous network analysis \cite{13}, but provides greater resolution and statistical certainty that these communities represent genuine constraints on recombination. However, the membership lists of communities derived from different HVRs matched each other for only 38\% of nodes on average. In addition to having widely varying community structures, HVR networks also differed from each other in component size and number of components illustrated in Figure \ref{fig4} and tabulated in Table \ref{table2}. Networks are visually illustrated in Figure \ref{fig4} and the number and size of components are given in Table \ref{table2}. Regardless of the number of communities detected, communities corresponded poorly to each other across HVRs. 

\subsubsection{Lack of consensus between HVR networks}
\begin{figure*}[t]
	\includegraphics[width=0.95\linewidth]{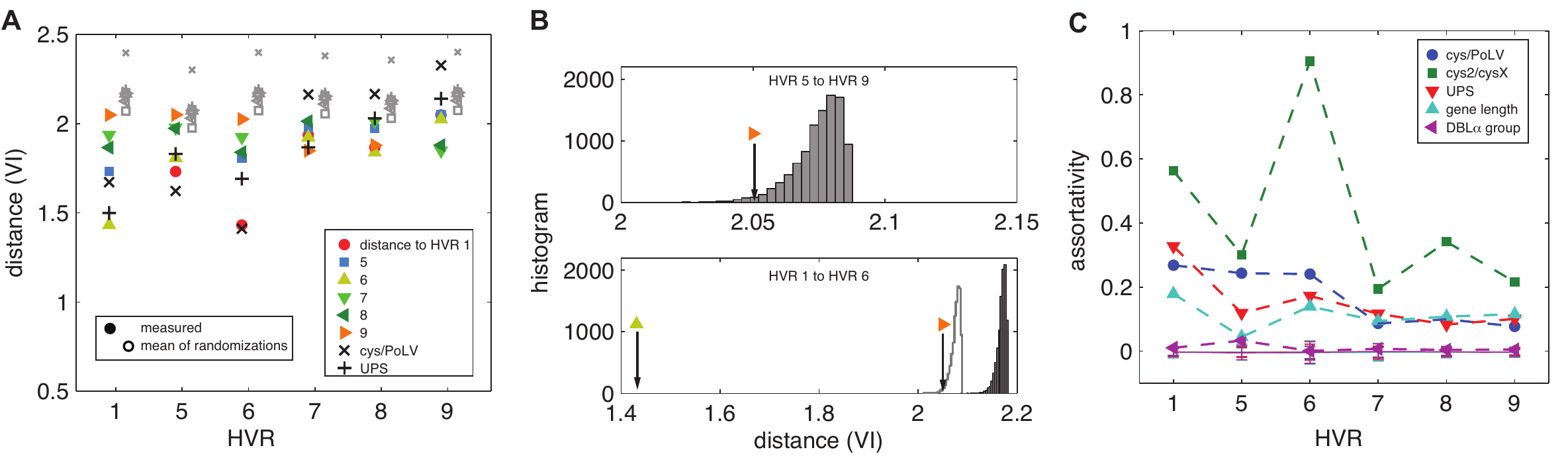}
	\caption{{\bf Community structures vary across HVRs.} (A) Variation of information (VI) measures the distance between two different partitions on the same set of nodes (two different sets of community assignments). For a given HVR we compute pairwise VI distances to recovered communities of other HVRs (colored symbols), cys/PoLV (+), UPS classifications (x), and to those in a null model (grey open symbols). Uncertainty in VI measurements is discussed in detail in Text S4. (B) HVRs 1 and 6 are close to each other, indicating that their communities strongly match each other. Histograms show the distributions of VI distances for one partition and 10,000 randomizations of the other. Top, the measured distance between HVRs 5 and 9 falls within the distribution of randomizations, indicated by the arrow. Bottom, the measured distance between HVRs 1 and 6 falls well outside the distribution of randomizations, indicated by the left arrow. For contrast, the silhouette of the top histogram is reproduced. (C) Most HVRs show moderate but positive levels of assortativity \cite{45}, the tendency for nodes with similar labels or values to be connected. Assortativity varies by HVR and by label (symbols). For all cases except DBL$\alpha$ classification (DBL$\alpha$0, DBL$\alpha$1, DBL$\alpha$2), assortativity was significantly higher than expected by chance. Solid lines with whiskers show mean assortativity $\pm$ one standard deviation for 10,000 randomizations of labels. Z-scores may be found in Figure \ref{figs5}A.}
	\label{fig6}
\end{figure*}

Traditional phylogenetic approaches often produce a consensus tree that reflects the most likely evolutionary trajectory of a particular gene. However, if patterns of recombination for two HVRs are relatively independent of each other, we expect the communities of one HVR network to match the communities of the other only to the extent they match by chance. In contrast, HVR networks with similar recombinational constraints will have common community structures. In order to quantify the distance between community assignments, we use the variation of information statistic \cite{40}, which is a distance metric on partitions. A small value indicates that two partitions are ÒcloseÓ to each other, such that the composition of one is highly correlated with the composition of the other. Figure \ref{fig6}A shows the pairwise distances between the inferred communities for $k=3$ and partitions defined by UPS and cys/PoLV classifications, across HVRs. (Plots for all values of k are in Figures \ref{figs4} and \ref{figs5}B.) 

In general, the communities of different HVRs were surprisingly dissimilar, except HVRs 1 and 6, and, to a lesser extent, HVRs 1 and 5. We compared the observed distances to an estimated distribution of pairwise distances for a null model in which we held one partitioning constant and computed distances for 10,000 random permutations of the other, for each pair, shown as grey symbols in Figure \ref{fig6}A and converted to z-scores in Figure \ref{figs5}B. Two examples of the randomized distributions are shown in detail in Figure \ref{fig6}B: the comparison of HVR 5 with HVR 9, and the comparison of HVR 1 with HVR 6. While the distance between HVRs 5 and 9 is smaller than the expected value of a random permutation (top subplot) it is significantly closer to its expected value than HVRs 1 and 6 (bottom subplot). We estimated statistical uncertainty in these measurements and found that in all cases, standard deviations were $\mathcal{O}(10^{-2})$, much smaller than the size of colored symbols plotted in Figure \ref{fig6}A. However, we note that this estimate is one of many possible measures of statistical uncertainty for network parameters, each of which is flawed in some way, which we discuss fully in Text S4. 

Although no pair of community assignments is farther apart than expected at random, most other community assignments are only very weakly similar to each other. The fact that individual HVRs feature clear community structure implies that there are evolutionary constraints on recombination; yet comparisons of community structure between HVR networks reveal only slightly more similarity than random, suggesting that recombinational constraints at different positions are almost completely independent of each other. Thus, variable selection pressures can be accommodated even within a single DBL$\alpha$ domain (distances are shown as a heatmap for all pair-wise comparisons in supplemental Figure \ref{figs4}). These patterns suggest that mosaic sequences behave as dynamic modules that can be shared among genes relatively intact, with conserved inter-mosaic regions acting as alignment guides in the recombination process. A key finding of this analysis, therefore, is that the relative independence of different HVR networks precludes the use of ÒconsensusÓ approaches sometimes used to combine trees; HVR networks were sufficiently different that they must be analyzed independently. 

\subsection{Relationship between HVR communities and previous classification systems}

In the absence of tools capable of handling extremely high rates of recombination, {\it var} genes have been variously classified by their domain structure and gene length, sequence characteristics, upstream promoter regions (UPS), position within the chromosome, and direction of transcription \cite{25,29,30,31,32,40,41}. However, the correspondence of these groups, which only partially overlap, remains ambiguous. As increasing volumes of {\it var} gene sequence data are produced from studies in the field, understanding how best to resolve and refine these approaches will be key to interpreting study outcomes. We compared previous classification systems, as well as each parasite genotype, to communities within each HVR network.

\subsubsection{Parasite {\it var} repertoires reflect population-level diversity}
Examining the sequences of individual parasites, we found that each of the seven parasitesÕ {\it var} sequences are evenly spread through the clusters of the network, rather than forming genome-specific communities (Figure \ref{figs1}), consistent with previous studies \cite{10}. Regardless of how many communities $k$ we choose, each parasite had at least one sequence in each recovered community, showing that a single parasite is not confined to an identifiable genotypic community but instead has samples of all major communities that we identified. This corroborates previous research showing that {\it var} genotypic diversity within a single parasite is as high as the diversity of the parasite population \cite{20}. This pattern is also consistent with theoretical work has shown that selective pressure on the {\it var} genes should create parasites with as wide a variety of genotypes as possible for immune evasion, while still preserving enough structure for adhesion and sequestration \cite{42}. Thus, each {\it P. falciparum} genome contains an antigenic repertoire that is effectively sampled from the diversity of the global pool of {\it var} genes. 

\subsubsection{Upstream promoter sequences and {\it var} gene length}
While HVR networks tend to differ from each other, their communities correspond to known upstream promoter sequence (UPS) groupings and {\it var} gene length. Upstream promoter sequences were previously categorized as UPS A to UPS E or Not Determined (ND) \cite{18}. DBL$\alpha$ with UPS D were not present in the 307 sequences examined. The inferred communities in HVRs 1 and 6-8 place nearly all UPS A sequences together, plotted in Figure \ref{fig7}A. The remaining communities comprise a mix of UPS B, C, and ND sequences. This implies that recombinant mixing of UPS A with B or C is comparatively rare, leading to the hypothesis that three ND sequences could be classified as UPSA: IT4var24, PF07\_0048, and IT4var51. Furthermore, we found no evidence for a strong separation of {\it var} genes into distinct groups for UPS B and C. Although UPS C genes are all found proximate to the centromeres of {\it P. falciparum} genomes, recombination appears to occur frequently between centromeric and subtelomeric UPS B genes, consistent with studies of chromosomal positioning during ectopic recombination \cite{14,43}, but not yet observed in vitro \cite{44}. 

Using the variation of information distance measure \cite{40}, Figure \ref{fig6}A shows that the community structures of HVR 1 and UPS communities are particularly close. This is reinforced by a measurement of assortative mixing by label \cite{45}, shown in Figure \ref{fig6}C, a measure of correlation among node labels that takes into account the connections of the network and may be computed without any community assignments. In particular, this reflects the fact that in HVR 1, many UPS A nodes tend to link almost exclusively to other UPS A nodes. Plots of all HVR networks colored by UPS group are found in Figure \ref{figs2}. We also found positive assortative mixing by {\it var} gene length (including all NTS, DBL, CIDR, and ATS domains) implying that genes of the similar length tend to link to each other, consistent with dynamical models of {\it var} gene evolution \cite{42} (Figure \ref{fig6}C). Thus, both UPS group and gene length may play roles in constraining recombination, suggesting that future models of evolution and recombination must reproduce these results.

\subsubsection{Cys2/Cys4 and PoLV classification}
\begin{figure*}
	\includegraphics[width=0.8\linewidth]{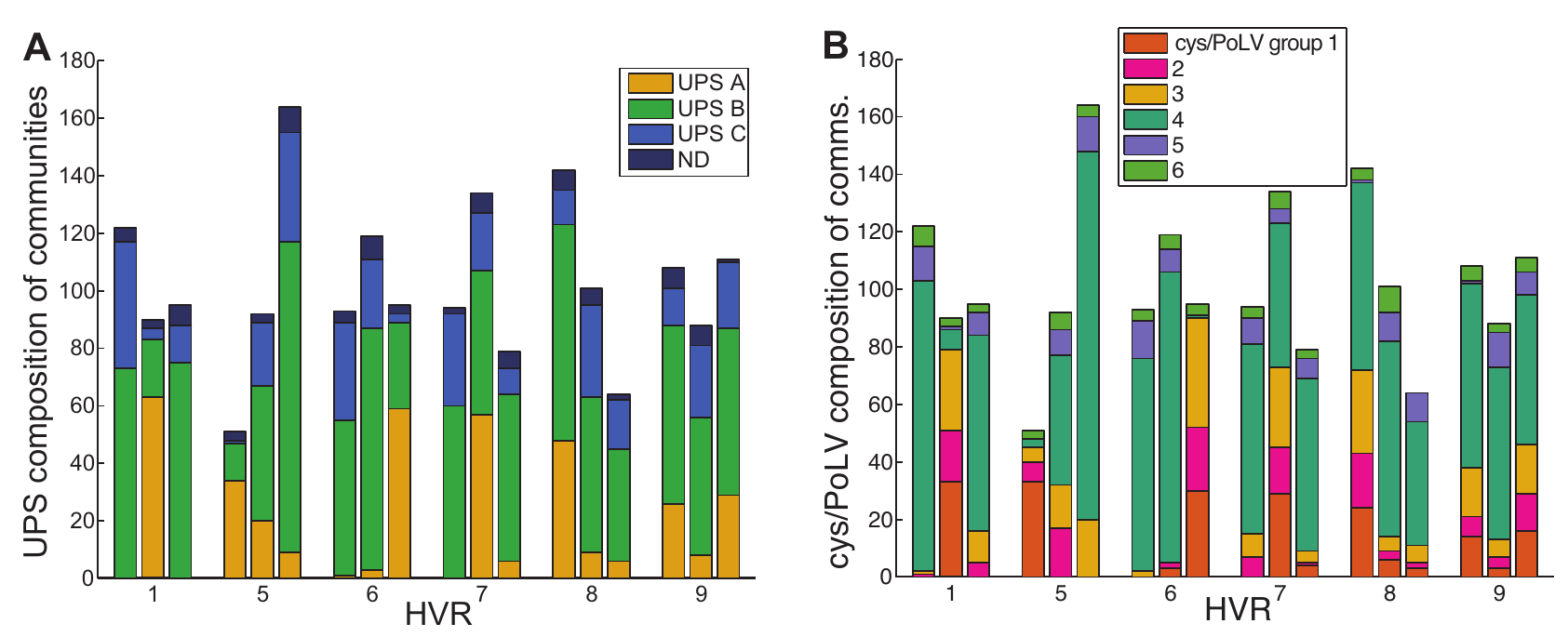}
	\caption{{\bf Correspondence of network communities to existing classifications.} Bars show the UPS group (A) and cys/PoLV group (B) composition of each of the three recovered communities. HVRs 1 and 6-8 show one community each in which a vast majority of UPS A sequences are found. HVR 6 shows one community in which a vast majority of group 1, 2, and 3 sequences are found, which are all characterized by having only two cysteines in HVRs 5 and 6.}
	\label{fig7}
\end{figure*}

Our method confirms and extends two previous analyses that identified patterns within the DBL$\alpha$ domain. In an alignment-free study of short tagged sequences corresponding to HVRs 5 and 6 from Kilifi, Kenya, it was found that sequences may be classified based on the number of cysteine residues present in HVR 6, and that more severe disease phenotypes were correlated with the presence of only two cysteines \cite{10,46}. These sequences are referred to as Òcys2Ó sequences, and others, most of which have four cysteines, are referred to as ÒcysX.Ó  The cys2/cysX classifications can be further classified according to a set of mutually exclusive sequences motifs called positions of limited variability (PoLV), splitting cys2/cysX categories into six cys/PoLV groups \cite{10,47}. Since the HVR network approach creates links based on shared block structure, it captures cys2/cysX  and cys/PoLV categorizations. Unsurprisingly, this is reflected most strongly in HVR6 (Figures \ref{fig6}A and \ref{fig6}C), but is also present in HVR1 whose nodes are strongly assortatively mixed by cys/PoLV group (Figure \ref{fig6}C), corroborating the result that HVRs 1 and 6 are structurally similar. There exists one community in HVRs 1, 6, and 5 that contains most of the cys2 sequences (Figure \ref{fig7}B). The correspondence of network structures with previous phenotype-associated sequences also suggests that the HVR network approach may be extended to map genotypic patterns to clinical phenotypes when applied to an appropriate data set that includes expression data and clinical information.

\subsubsection{DBL$\alpha$ subdomain classification}
The DBL$\alpha$ domains we use here were previously analyzed and categorized extensively using tree-based methods, revealing many new subclassifications of the previously identified DBL$\alpha$0 and DBL$\alpha$1 \cite{41,43} as well as a new classification, DBL$\alpha$2 \cite{18}. We examined HVR networks for evidence of strong associations between network community structure and subclassifications, but found that only DBL$\alpha$1.3 sequences had a strong tendency to link to other sequences from the same subclassification, particularly in HVRs 7, 8, and 9, where they formed cliques on the periphery of the networks. While the sequence differences of the DBL$\alpha$1.3 subclass have already been noted \cite{18}, we also find that the DBL$\alpha$0.3 subclass tended to link to each other in HVRs 1 and 8, though not in a clique. No strong correlations were found between network structures and the broader classes of DBL$\alpha$0, DBL$\alpha$1, and DBL$\alpha$2, and networks were not strongly assortative by DBL$\alpha$ class, as shown in Figure \ref{fig6}C. This is perhaps unsurprising, given the different focus of our method, but it highlights the fact that any classification of recombinant genes that is based on a tree-like phylogeny may not be informative about the process of recombination. 

\subsection{{\it var} gene micromodularity facilitates diversity generation without loss of function}

\begin{figure}[b]
	\includegraphics[width=0.95\linewidth]{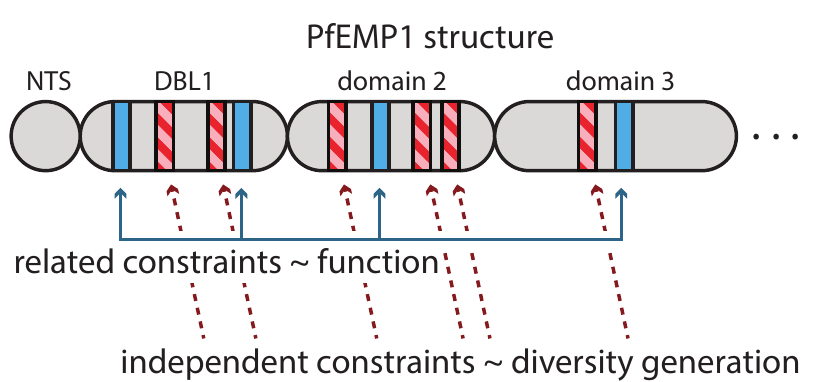}
	\caption{{\bf Schematic of HVR hypothesis.} Each HVR of DBL$\alpha$ is highly recombinant, yet recombinational constraints, as revealed by network community structure, vary by HVR. We find that some are strongly related (HVRs 1 and 6) while others are independent (HVRs 5 and 9). This suggests that those HVRs that diversify and recombine under constraints that are independent of each other may have the primary purpose of immune evasion by diversity generation (red stripes). Those HVRs whose recombinational constraints are related to each other may recombine only under functional constraints (blue). Independent or related constraints may also be found in other domains throughout the larger PfEMP1 molecule.}
	\label{fig8}
\end{figure}

Our approach highlights and analyzes the within-domain modularity of DBL$\alpha$. {\it var} genes are by definition modular, with variable numbers and types of DBL and CIDR domains clustered into larger groups that primarily correspond to UPS A and UPS B or C. We have shown that in addition to modular domains that can be shuffled between loci, there are also relatively regularly spaced modular mosaics within the DBL$\alpha$ domain that are shuffled, under constraints, between {\it var} genes. Previous studies have suggested that the conserved blocks in DBL domains may provide structural support for the protein while the variable regions in between are loops in the protein designed specifically for antigenic variation under diversifying selection \cite{17,33}. Here we offer a more nuanced hypothesis, shown schematically in Figure \ref{fig8}: HVRs under related recombinational constraints may have important functional roles in the PfEMP1 molecule, while other HVRs may exist primarily for purposes of antigenic variation. 

We find that HVRs 1, 6, and to a lesser extent 5, have similar non-random network community structure that corresponds strongly with structural amino acid residues and classification systems with known associations with severe disease \cite{47}. These regions of the domain may therefore be functionally constrained and play a specific role in binding. Interestingly, HVRs 5 and 6 correspond to the short tag sequences that have previously been amplified from field isolates. If our hypothesis is correct, it would explain why the clustering of these sequence tags exhibit meaningful associations with disease outcome \cite{46,48}. The remaining HVRs 7-9, have highly heterogeneous community structures that bear almost no relation to each other, suggesting that their primary role is in the generation of diversity for the purpose of immune evasion. 

Having both correlated and uncorrelated recombinational constraints across multiple HVRs thus provides at least two important evolutionary benefits to the parasite: i) individual mosaics that are functionally important can retain their function without compromising the generation of diversity across the rest of the domain, and ii) recombination could produce variants through different combinations of modules more rapidly than mutation or random recombination, and without risking the recombinational equivalent of error catastrophe that occurs in systems with very high mutation rates \cite{49}. In other words, the parasite population may be rapidly shuffling ancient sequence mosaics into new combinations across some HVRs, while also preventing the degeneration of structurally important regions of the protein that are involved in binding. The population of genes may thereby balance a need for new diversity with functional requirements.
 
The varying correspondence of HVR communities with previously defined {\it var} gene groupings implies that the different classification schemes complement each other, providing insights into different aspects of {\it var} gene evolution, likely representing nested or hierarchical recombinant clusters. We measure the extent to which previously defined groupings are reflected in the links of our networks using assortativity, shown in Figure \ref{fig6}C. Importantly, the DBL$\alpha$ group assortativity is much lower than all the others, demonstrating further that while there are clear structures in HVR networks, they are not the same as the classifications based on trees. Such tree-based classifications explain the development of large-scale structure over the relatively longer timescales of mutation, whereas communities detected within recombination networks here shed light on functional or even mechanistic constraints on recombination occurring more recently. 

\section{Conclusions}

The method presented here can accurately extract block-sharing networks from the most highly recombinant regions of protein sequences. While a multiple-alignment is used to identify HVRs (Figure \ref{fig1}A), we find that HVR boundaries are robust to changes in alignment parameters, and the remaining steps of our network extraction process (Figures \ref{fig1}B-D) are entirely alignment-free. Since the structure of such networks reveals patterns of recombination, strong communities within networks are indicative of functional or evolutionary constraints on recombination within the underlying population. By applying this technique to the DBL$\alpha$ domain of {\it var} genes we find that different locations of the domain produce different communities. Were the constraints identical in each domain, we would expect network communities to be similar to each other and exhibit small variation of information distance. The fact that network communities differ therefore indicates that while constraints exist at each location in the domain, they also vary by location. We suggest that this lack of correlation between constraints allows the DBL$\alpha$ domain to possess exponentially more complexity while simultaneously remaining functional, avoiding recombinant error catastrophe despite extremely high rates of recombination.  

The combination of principled and alignment-free network construction methods with state-of-the-art generative models for community detection may open the door to new research areas, linking evolution, function, and clinical phenotypes in a range of genetically diverse pathogens. While we demonstrate our method using DBL$\alpha$, it could also be applied to other highly recombinant {\it var} domains, other {\it P. falciparum} genes such as the rif and cirs families, or other pathogens, such as HIV, the pneumococcus, or trypanosomes. Methodologically, extensions include the development and application of stochastic block model community detection in weighted networks, treating the separate HVR networks as a single multiplex network \cite{50}, and generalizing the current process to inexact sequence matches.

\section{Acknowledgements}
We thank Nicholas J. Croucher for assistance in making phylogenetic trees, and Peter C. Bull and Mario Recker for helpful discussions.

\clearpage
\renewcommand{\thefigure}{S\arabic{figure}}
\setcounter{figure}{0}
\renewcommand{\thetable}{S\arabic{table}}
\setcounter{table}{0}
\renewcommand{\theequation}{S\arabic{equation}}
\setcounter{equation}{0}

\begin{figure*}
	\includegraphics[width=0.95\linewidth]{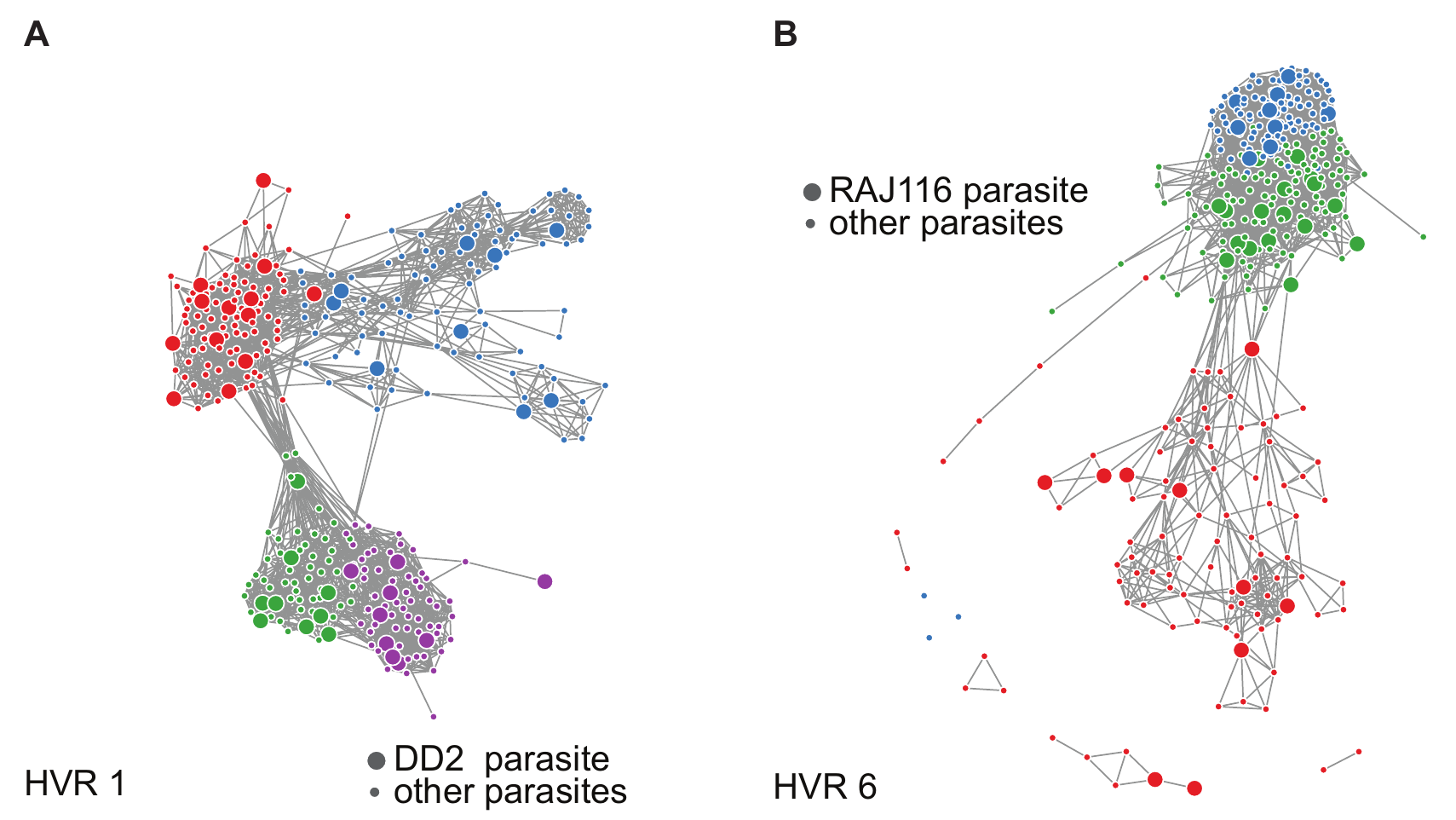}
	\caption{{\bf Diversity of individual parasites.} Sequences from a single parasite (large nodes) are distributed evenly throughout the recovered groups. Thus, the diversity within a single parasite is comparable to the diversity of the entire network. (A) HVR 1 with four inferred communities, highlighting the DD2 parasite, and (B) HVR 6 with three inferred communities, highlighting the RAJ116 parasite provide two representative examples. Figures depict differing numbers of communities to show that single parasite genomes are distributed throughout network communities, even as the number of communities increases. An interactive version of this figure may be found at \href{http://danlarremore.com/var}{http://danlarremore.com/var}. Networks are displayed using a force-directed algorithm that allows a system of repelling point-charges (nodes) and linear springs (links) to relax to a low-energy two dimensional configuration, allowing for visualization of network communities.}
	\label{figs1}
\end{figure*}

\begin{figure*}
	\includegraphics[width=0.8\linewidth]{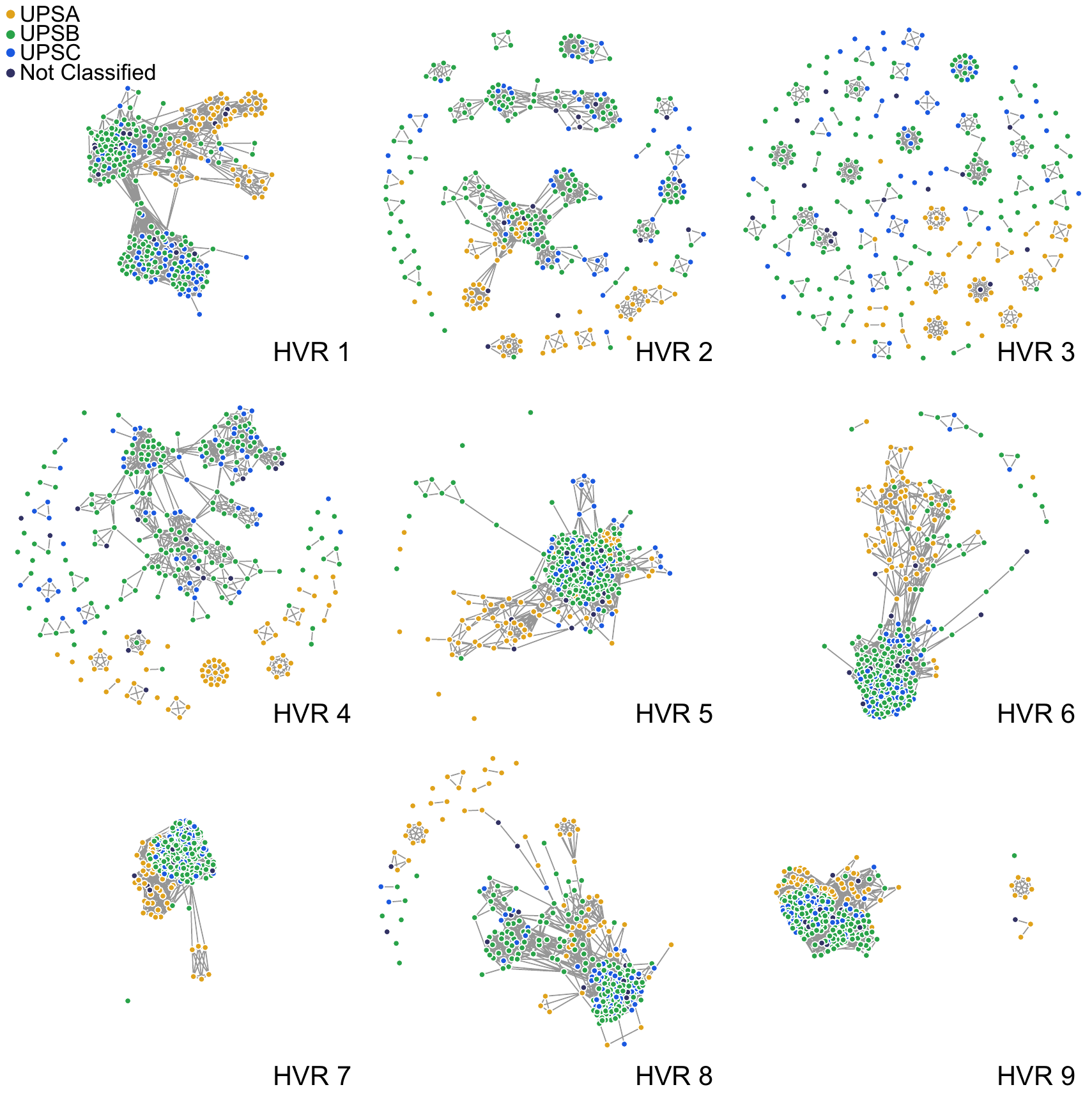}
	\caption{{\bf Nine HVR networks colored by UPS grouping.} DBL$\alpha$ HVR networks are colored by Upstream Promoter Region (UPS) with categories of A,B,C, or Not Determined. An interactive version of this figure with varying communities and node labels may be found at \href{http://danlarremore.com/var}{http://danlarremore.com/var}. Networks are displayed using a force-directed algorithm that allows a system of repelling point-charges (nodes) and linear springs (links) to relax to a low-energy two dimensional configuration, allowing for visualization of network communities.}
	\label{figs2}
\end{figure*}

\begin{figure*}
	\includegraphics[width=0.95\linewidth]{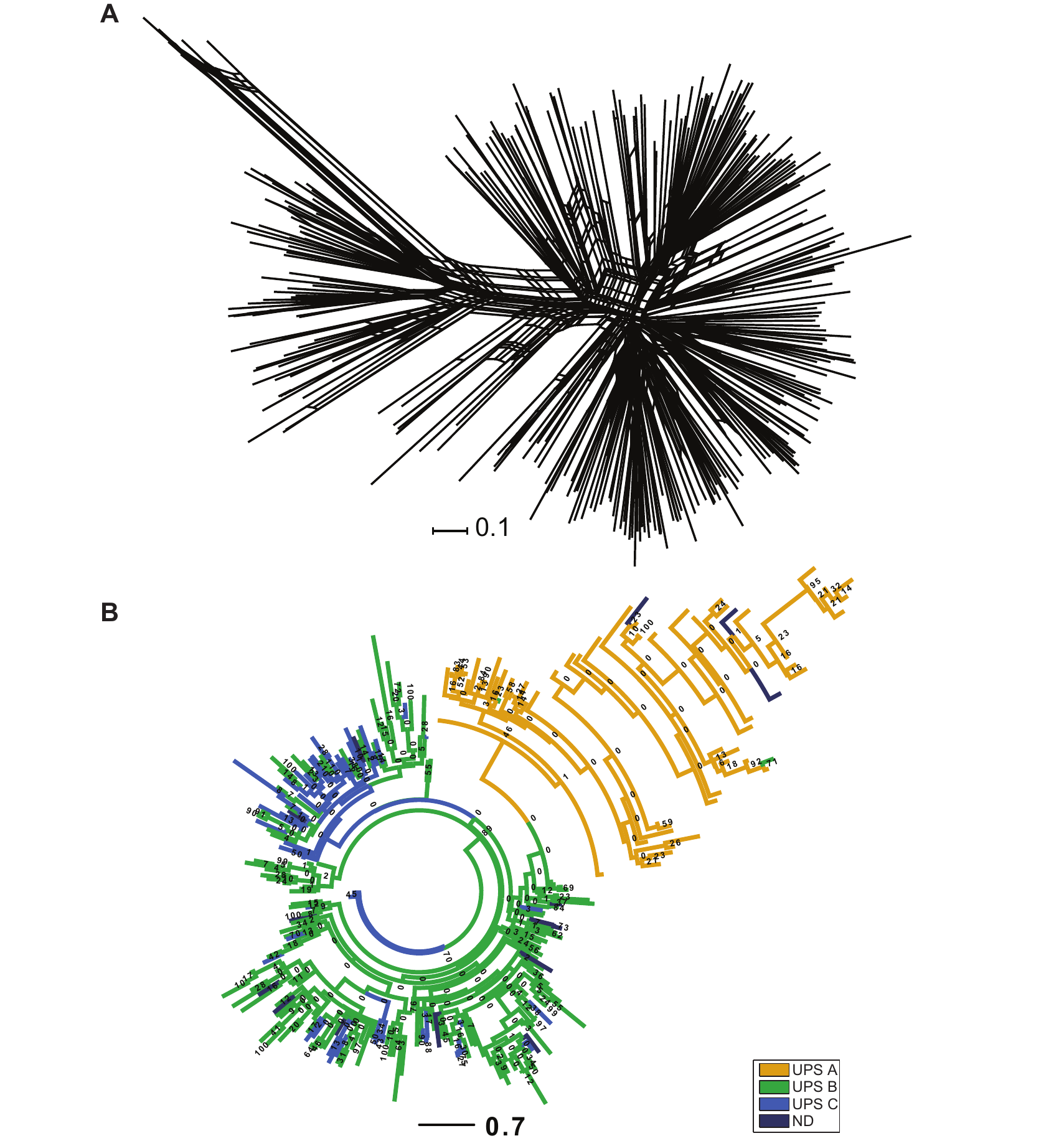}
	\caption{{\bf Phylogenetic trees of DBL$\alpha$ sequences. } (A) A SplitsTree diagram displays a phylogenetic tree in which ambiguous or recombinant sequences are linked with a cross-link, as described in S3. The presence of many cross links and very little reliable tree structure demonstrates the need for tools beyond traditional phylogenetic trees and networks. (B) A phylogenetic tree was constructed using RaxML 7.0 as described in S3, and color-coded by UPS group. While the tree classifies nearly all UPSA sequences together, poor bootstrap values indicate that this phylogeny has low statistical reliability. }
	\label{figs3}
\end{figure*}

\begin{figure}
	\includegraphics[width=0.95\linewidth]{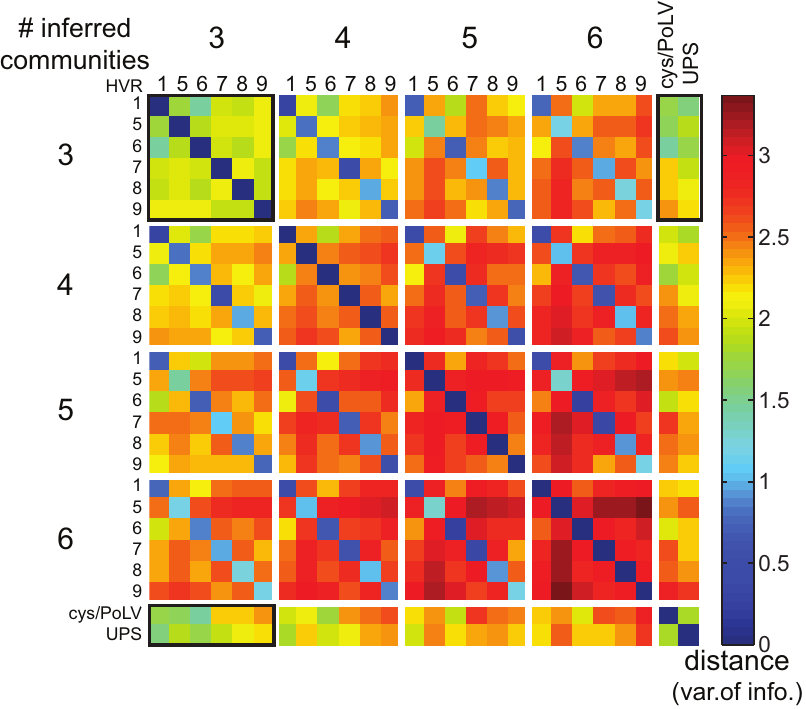}
	\caption{{\bf Pairwise distances between HVR communities. } Similarities in community structures are measured here by the variation of information (VI) metric [40],  pairwise for each HVR and for k=3, 4, 5, and 6 communities, as well as cys/PoLV and UPS classifications. Small distances on the ±6-, ±12-, and ±18-diagonal stripes indicate that each HVR is more similar to itself (regardless of the number of communities inferred) than it is to any other HVR. This indicates that HVR networks do possess distinct community structures, but that such structures vary widely by HVR.  A similar figure showing z-scores of VI from randomization of labels is included in Figure \ref{figs5}B. VI values from the regions in the black boxes are displayed graphically in Figure \ref{fig6}A. Accordingly, HVRs 1 and 6 are much closer than any other pair of partitions of two different HVRs for a variety of k values.}
	\label{figs4}
\end{figure}

\begin{figure*}
	\includegraphics[width=0.95\linewidth]{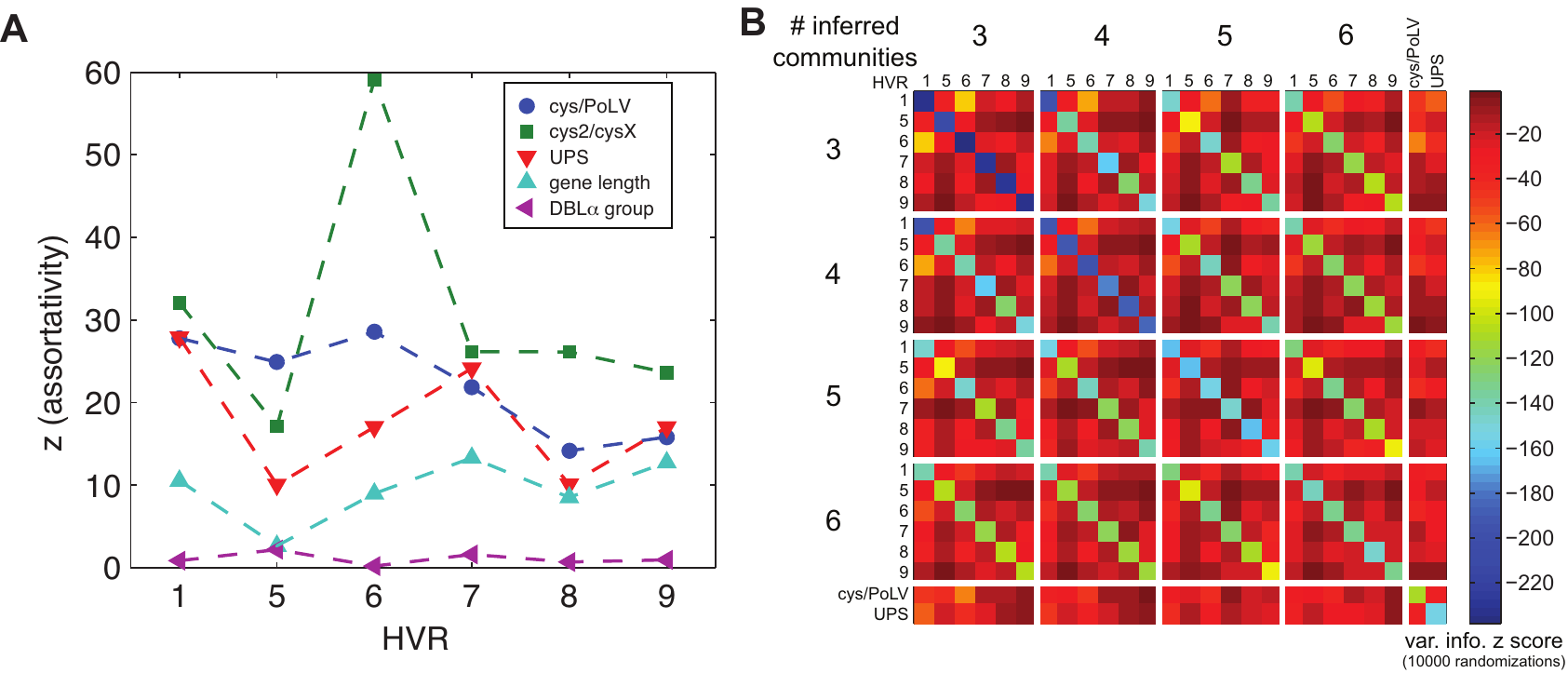}
	\caption{{\bf Node label randomization tests. } Results from randomization trials show that the assortativities shown in Figure \ref{fig6}C and some of the distances shown in Figure \ref{figs4} are statistically significant, because the probability that the null model would yield as small or smaller distances is minute. (A) Each point is an assortativity z-score calculated from 10,000 randomizations of a network's node labels, showing that network assortativities by label and gene length reflect a preference of sequences to recombine with other similar sequences. (B) Each point is a VI z-score from 10,000 randomizations of node labels. As in Figure \ref{figs4}, HVRs are more similar to themselves (with any number of communities) than they are to each other. HVRs 1 and 6 are again shown as more similar to each other than other pairs, followed closely by HVR1 and UPS, and HVR6 and cys/PoLV communities. It is important to note, however, that randomized variation of information distances are not normally distributed, and so these z-scores should not be interpreted as such.}
	\label{figs5}
\end{figure*}

\begin{figure*}
	\includegraphics[width=0.95\linewidth]{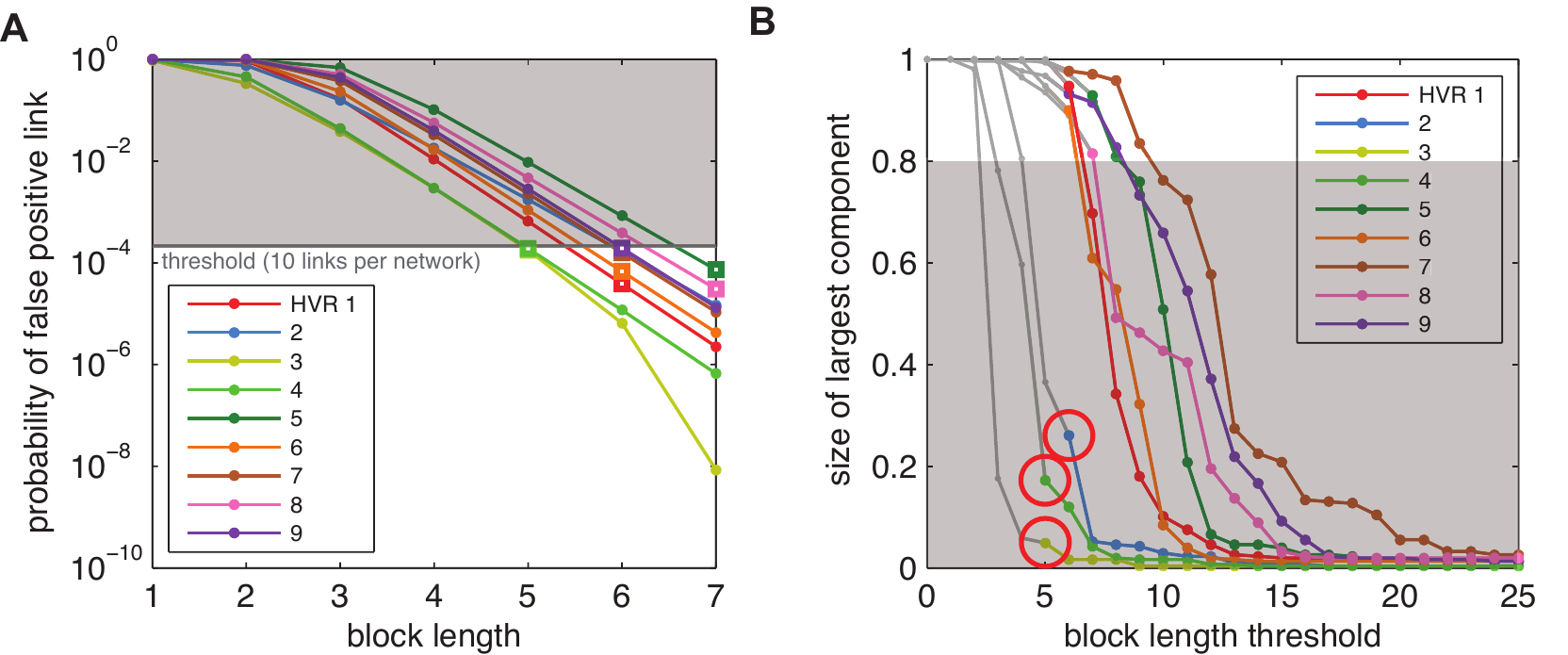}
	\caption{{\bf Choice of link noise threshold (Figure \ref{fig2} with all HVRs shown).} While Figures \ref{fig2}A and \ref{fig2}C illustrate the method by which link noise threshold is selected, we excluded other HVRs to provide a more legible and understandable figure. Here we show all nine HVRs. }
	\label{figs6}
\end{figure*}

\begin{widetext}
\section{Text S1: Null model for sharing of protein sequence substrings}

Creating networks requires comparing many pairs of amino acid sequences. For each pair, we seek the length of the longest substring shared by both sequences. Some sequence pairs may have zero amino acids in common, in which case the length of the longest shared substring will be zero. However, most pairs of sequences will share a common substring that is at least one amino acid in length. How long must a shared substring be in order to be retained as a meaningful link in the network? Here, we compute the probability that a shared substring of a particular length arises simply by chance in order to allow a principled method by which to choose a threshold for link retention. 

Suppose that we construct two sequences called $n$ and $m$, of length $|n|=N$ and $|m|=M$, respectively, by drawing amino acids one-by-one from a replenishing urn from which each acid $i$ will be drawn with probability $p_i$.  What is the probability that they share one or more substrings of length $L$? 

For convenience, we define $\mathcal{S}^{L} \equiv$ the set of all possible amino acid sequences of length $L$. Since there are 20 amino acids derived from nucleotide codons, $| \mathcal{S}^{L} |=20^{L}$. We enumerate these by $i$, so that $ \mathcal{S}^{L} = \{s_i \}_{i=1}^{20^{L}}$

\begin{align}
	\text{Pr(at least one shared string of length }L) &= 1 - \text{Pr(no shared strings of length }L) \nonumber \\
	&= 1 - \prod_{s_i \in \mathcal{S}^{L}} \text{Pr(string }s_i\text{ is not shared)} \nonumber\\
	&= 1 - \prod_{s_i \in \mathcal{S}^{L}} 1 - \text{Pr(string }s_i\text{ is shared at least once)} \nonumber\\
	&= 1 - \prod_{s_i \in \mathcal{S}^{L}} 1 - \text{Pr(string }s_i\text{ is in sequence } n \text{ AND in sequence } m \text{)} \nonumber\\
	&= 1 - \prod_{s_i \in \mathcal{S}^{L}} 1 - \text{Pr(string }s_i\text{ is in sequence } n \text{)}\times \text{Pr(string }s_i\text{ is in sequence } m \text{)}
	\label{eq-shared}
\end{align}
We pause here for a side calculation, noting that there are $N-L+1$ different ways to position a string of length $L$ in a sequence of length $N$:
\begin{align}
	\text{Pr(string }s_i\text{ is in sequence }n) &= \text{1- Pr(string }s_i\text{ is not in sequence }n\text{ anywhere)} \nonumber \\
	\displaystyle &= 1- \prod_{j=1}^{N-L+1}[1-\text{Pr(string }s_i\text{is found in position $j$)}]\nonumber\\
	\displaystyle &= 1- [1-\text{Pr(string }s_i\text{is drawn from urn)}]^{N-L+1}
	\label{eq-sharedsidecalculation}
\end{align}
with the last step coming from the assumption of homogeneity and independence of the probabilities at different positions of the sequence, and therefore all $N-L+1$ positions may be treated equally as being drawn from an urn IID. This assumption of independence means that we will overestimate the probability of sequence sharing. Therefore, after this point, the calculation yields an upper bound for probability. Substituting Eq.~\eqref{eq-sharedsidecalculation} into Eq.~\eqref{eq-shared} yields 
\begin{equation}
	\displaystyle\text{Pr(at least one shared string of length }L) \leq 1 - \prod_{s_i \in \mathcal{S}^{L}} \bigg \{1 - \bigg [1- (1-\text{Pr}(s_i))^{N-L+1} \bigg ] \times \bigg [1- (1-\text{Pr}(s_i))^{M-L+1} \bigg] \bigg \}.
		\label{diffn}
 \end{equation}

When calculating Pr($s_i$) in practice we model strings as being drawn $i.i.d.$ from a multinomial distribution in which each of the amino acids $i$ has probability $p_i$, which we set to the maximum likelihood value $f_i / T$, where $f_i$ is the empirical frequency of acid $i$ and $T$ is the total number of acids observed. Since amino acid composition varies with HVR, this approach takes into account the composition of each HVR separately. Similarly, the lengths of sequences in each HVR are different as shown in Table \ref{table1}. To simplify computation, we let $N=M$ in Eq.~\eqref{diffn}, yielding

\begin{equation}
	\displaystyle\text{Pr(at least one shared string of length L)} \leq 1 - \prod_{s_i \in \mathcal{S}^{L}} \bigg \{1 - \bigg [1- (1-\text{Pr}(s_i))^{N-L+1} \bigg ]^2  \bigg \}.
	\label{samen}
 \end{equation}

We use Eq. \eqref{samen} to compute each curve in Figure \ref{fig2}A, setting $N$ to the median length of each HVR (Table \ref{table1}), allowing a translation between an expected number of false positives and a length threshold below which links should be discarded. Since it is an upper bound, strictly adhering to the threshold that it prescribes will result in fewer false positive links, but may also result in fewer true positive links.
\end{widetext}

 \section{Text S2: Validation on synthetic data}

In order to demonstrate that the community structures found in real networks may correspond to real constraints on recombination, we create synthetic amino acid sequences and allow them to recombine under various levels of a recombination constraint and measure the accuracy of our method. The specific procedures followed were: (i) Create sequences: choose an empirically measured distribution of amino acid frequencies from an existing HVR. Using the empirical distribution as the parameters of a multinomial distribution, draw amino acids IID in order to generate 60 sequences, each of length 30. Arbitrarily separate the sequences into three groups of 20 sequences each, labeled $A$, $B$, and $C$. (ii) Simulate recombination: uniformly at random, choose an existing sequence as mother, noting her group. With probability $p$, choose a father uniformly at random from the same group as the mother; with probability $(1-p)$ choose the father uniformly at random from all sequences. Mimic a gene conversion by choosing uniformly at random a block of length $a$ from the mother and length $b$ from the father, creating a child that is a copy of the mother, but with the father's block replacing the mother's. Here we choose $a,b \sim$Unif$[8,12]$. Place the child into the same group as the mother. Repeat this step 1000 times, each time placing the child into the pool of sequences from which mother and father may be selected, resulting in a total of 1060 sequences. Discard the original 60 progenitor sequences. (iii) Create networks. Forgo the step of identifying HVRs---we assume that our synthetic data comes from a single HVR. (iv) Find communities using the degree-corrected stochastic blockmodel with $k=3$ communities, classifying nodes as community $X$, $Y$, or $Z$. (v) Measure accuracy as the fraction of nodes that are matched between $X$, $Y$, $Z$, and $A$, $B$, $C$. There are six possible pairing sets of the original communities and the detected communities, so we measure accuracy as the pairing set that results in the maximal value. This means that simply guessing communities uniformly at random would result in accuracies slightly higher than $1/3$ because the accuracy-maximizing pairing set is bounded from below by $1/3$ and is often larger, stochastically. (vi) Measure accuracy for 25 replicates at various values of $p$. We note that when $p=0$ accuracy is still slightly better than random guessing due to weak network structures induced by heredity patterns. 

 \section{Text S3: Phylogenetic analyses}
 
Protein sequences were aligned using MUSCLE v3.8 (PMID: 15034147 \cite{Edgar:2004bo}). Different models of sequence evolution were then evaluated using ProtTest v3.2 (PMID: 21335321 \cite{Darriba:2011ht}); based on the Bayesian information criterion, this found the WAG matrix combined with a gamma rate distribution and a calculation for the proportion of invariant sites to be the most appropriate evolutionary model. These settings were therefore used to generate a maximum likelihood distance matrix and phylogeny using RAxML v7.0 (PMID: 16928733 \cite{Stamatakis:2006de}), with 100 bootstrap replicates produced to measure support for internal branches. The distance matrix was analysed using SplitsTree v4 (PMID: 9520503 \cite{Huson:1998um}) as a further measure of uncertainty in the phylogeny.

\section{Text S4: Uncertainty in measurements of VI from inferred partitions}

In the main text, a stochastic block model is fit to network data, generating a partition of the network nodes into communities, for each HVR. The distance between partitions, as measured by variation of information (VI), provides an idea of how similar or different community structures are between HVRs. The question of whether or not the difference between certain VI distances is large or small is an important one. If the differences are large, then we can say the pair of partitions really are different from each other, while if the difference is small, we conclude the opposite. However, inference of the parameters of the stochastic block model raises two questions about uncertainty: 
\begin{enumerate}
	\item What is the uncertainty in VI due to uncertainty in the likelihood maximization algorithm?
	\item What is the uncertainty in VI due to uncertainty inherent to the likelihood function itself?
\end{enumerate}
The first question is straightforward to address, yet the second is less so; uncertainty of likelihood functions of stochastic block models is the subject of ongoing theoretical research. This supplement provides a more thorough and lengthy discussion of the technicalities involved in answering both. 

\subsubsection{1. What is the uncertainty in VI due to uncertainty in the likelihood maximization algorithm?}

The algorithm used to find the maximum likelihood partition is described fully by Karrer and Newman and is similar to the Kernighan-Lin partitioning algorithm: 
\begin{quote}
Briefly, in this algorithm we divide the network into some initial set of K communities at random. Then we repeatedly move a vertex from one group to another, selecting at each step the move that will most increase the objective functionÑ or least decrease it if no increase is possibleÑsubject to the restriction that each vertex may be moved only once. When all vertices have been moved, we inspect the states through which the system passed from start to end of the procedure, select the one with the highest objective score, and use this state as the starting point for a new iteration of the same procedure. When a complete such iteration passes without any increase in the objective function, the algorithm ends. As with many deterministic algorithms, we have found it helpful to run the calculation with several different random initial conditions and take the best result over all runs.
\end{quote}
Since we initialize this algorithm with a random initial condition, due to roughness of the optimization landscape, a different local maximum may be found each time the algorithm is run. We can therefore calculate uncertainty in VI by the following procedure.
\begin{enumerate}
	\item Rerun the algorithm M times, computing an ensemble of M optimal partitions and their likelihoods.
	\item Compute VI pairwise between all pairs of optimal partitions, yielding M2 estimates.
	\item Compute a weighted average and weighted standard deviation of the ensemble where the weight of each datum is the sum of the two partitionsÕ log likelihoods. 
\end{enumerate}
This procedure is similar to a Bayesian model averaging approach, and allows for a statistically rigorous estimate of the underlying variance of the point estimates of the VI between each pair of HVRs. When we performed this analysis on our data, we found that standard deviations of VI distances were all O(10-2), suggesting that the roughness of the optimization landscape has little effect on the VI distances shown in Figure \ref{fig6}.

\subsubsection{2. What is the uncertainty in VI due to uncertainty inherent to the likelihood function itself?}

When data are scalar- or vector-valued this question can be answered in several ways that will provide reliable and statistically principled answers. One is to perform a bootstrap of the original data in order to estimate the bootstrap distribution of the estimated parameter. Another is to examine the local curvature of the likelihood function around the maximum likelihood estimate (MLE). Asymptotically, these approaches converge on the same uncertainty estimates.

However, the data here are not scalar- or vector-valued, but are instead relational, i.e., a network, and their interdependence violates the underlying assumptions of the bootstrap that would otherwise allow it to produce statistically rigorous uncertainty estimates. As a result, a bootstrap on network nodes does not produce an answer to question 2 because bootstrap networks are statistically dissimilar to the empirical data in crucial ways. 

For example, a bootstrapped network, whose nodes are resampled from original data with replacement, will have sampled some nodes multiple times and other nodes zero times. Those nodes that are sampled multiple times will form identical cliques with their clones, while nodes that have been sampled zero times may fragment the network into more components than in the original network. Resampled networks tend to have higher clustering coefficients (density of triangles) and occasionally more components as a result. Notably, the goal of this step of uncertainty analysis is to generate a distribution of plausible replicates of the test statistic, so if the resampled data sets are structurally very different from the empirical one, then the estimated distribution is not informative of the uncertainty in the statistic calculated from the original data set. 

This may be demonstrated in the case of VI by creating two realizations of the generative stochastic block model, producing two networks with known community structure. Inference of communities using the stochastic block model will produce extremely similar (though not identical, due to noise) partitions, and thus VI will be very small. However, bootstrapping VI in this case will result in a distribution that is comparatively very wide and is also not centered at the empirical value. 

Alternatively, one could conceive of bootstrapping VI by resampling edges instead of nodes, yet this, too, does not solve the problem. Edges in recombination networks represent characteristics of the vertices and thus are not independent of each other. Furthermore, the SBM used here is not defined for multi-graphs, which resampling the edges would produce.
 
Both of these issues point to a fundamental issue with bootstrapping network statistics: individual vertices are not IID draws of some underlying distribution, as in the case for most scalar- and vector-valued data, that can be considered independently of each otherÑa requirement for the bootstrap to produce consistent estimates. Instead, our networks are highly structured sets of interrelationships. 

One may also consider a leave-k-out technique (e.g., a statistical jackknife) in which, rather than resampling with replacement as in bootstrapping, k sequences are left out. This process has the major advantage of avoiding some of the non-random structural perturbations discussed above. Unfortunately, it has two large disadvantages that make it a non-viable alternative for examining the VI statistic. First, reducing the number of sequences used to construct the network changes the scale on which the VI is calculatedÑVI is bounded between 0 and roughly the log of the number of verticesÑwhich makes these VIs non-comparable to the empirical value. Second, subsampling presents similar violations of independence assumptions as in the bootstrap, and as a result the nice statistical properties (under IID conditions) of subsampling are not guaranteed.

Rather than a resampling approach, one might attempt to directly measure the local curvature of the likelihood function by Taylor expanding about the MLE. However, this requires a nice analytic form that may be expanded, as well as a likelihood function that is convex, i.e. has a single global maximum. The stochastic block model meets neither of these criteria. A more principled solution may be to sample the local optima of the likelihood function and estimate the variance in the test statistic using this ensemble (as described in the answer to question 1 above), yet this technique may confound likelihood function uncertainty with the uncertainty of the search algorithm.

\medskip

The conclusion is that uncertainty, as well as model selection, for stochastic block models (and principled inference techniques for networks in general) are not yet well studied enough to provide a definite answer to question 2. The answer to question 1 is optimistic, but we have included the preceding discussion so that readers interested in applying the techniques described here may avoid statistical pitfalls in their work. 


\begin{thebibliography}{99}
\bibitem{1} World Malaria Report (2012) World Malaria Report. World Health Organization: 1Ð33.

\bibitem{2} Bull PC, Lowe BS, Kortok M, Molyneux CS, Newbold CI, et al. (1998) Parasite antigens on the infected red cell surface are targets for naturally acquired immunity to malaria. Nature medicine 4: 358Ð360.

\bibitem{3} Bull PC, Marsh K (2002) The role of antibodies to Plasmodium falciparum-infected-erythrocyte surface antigens in naturally acquired immunity to malaria. Trends in Microbiology 10: 55Ð58.

\bibitem{4} Dodoo D, Staalsoe T, Giha H, Kurtzhals JA, Akanmori BD, et al. (2001) Antibodies to variant antigens on the surfaces of infected erythrocytes are associated with protection from malaria in Ghanaian children. Infection and Immunity 69: 3713Ð3718. doi:10.1128/IAI.69.6.3713Ð3718.2001.

\bibitem{5} Giha HA, Staalsoe T, Dodoo D, Roper C, Satti GM, et al. (2000) Antibodies to variable Plasmodium falciparum-infected erythrocyte surface antigens are associated with protection from novel malaria infections. Immunology letters 71: 117Ð126.

\bibitem{6} Newbold CI, Pinches R, Roberts DJ, Marsh K (1992)  Plasmodium falciparum: The human agglutinating antibody response to the infected red cell surface is predominantly variant specific. Experimental parasitology 75: 281Ð292.

\bibitem{7}	Chan J-A, Howell KB, Reiling L, Ataide R, Mackintosh CL, et al. (2012) Targets of antibodies against Plasmodium falciparumÐinfected erythrocytes in malaria immunity. J Clin Invest 122: 3227Ð3238. doi:10.1172/JCI62182DS1.

\bibitem{8}	Bull PC, Lowe BS, Kortok M, Marsh K (1999) Antibody recognition of Plasmodium falciparum erythrocyte surface antigens in Kenya: evidence for rare and prevalent variants. Infection and Immunity 67: 733Ð739.

\bibitem{9}	Kyes S, Horrocks P, Newbold C (2001) Antigenic variation at the infected red cell surface in malaria. Annual Reviews in Microbiology 55: 673Ð707.

\bibitem{10}	Bull PC, Berriman M, Kyes S, Quail MA, Hall N, et al. (2005) Plasmodium falciparum Variant Surface Antigen Expression Patterns during Malaria. PLoS Pathog 1: e26. doi:10.1371/journal.ppat.0010026.st003.

\bibitem{11}	Frank M, Kirkman L, Costantini D, Sanyal S, Lavazec C, et al. (2008) Frequent recombination events generate diversity within the multi-copy variant antigen gene families of Plasmodium falciparum. International Journal for Parasitology 38: 1099Ð1109. doi:10.1016/j.ijpara.2008.01.010.

\bibitem{12}	Atkinson HJ, Morris JH, Ferrin TE, Babbitt PC (2009) Using Sequence Similarity Networks for Visualization of Relationships Across Diverse Protein Superfamilies. PLoS ONE 4: e4345. doi:10.1371/journal.pone.0004345.t002.

\bibitem{13}	Bull PC, Buckee CO, Kyes S, Kortok MM, Thathy V, et al. (2008) Plasmodium falciparumantigenic variation. Mapping mosaic vargene sequences onto a network of shared, highly polymorphic sequence blocks. Molecular Microbiology 68: 1519Ð1534. doi:10.1111/j.1365-2958.2008.06248.x.

\bibitem{14}	Freitas-Junior LH, Bottius E, Pirrit LA, Deitsch KW, Scheidig C, et al. (2000) Frequent ectopic recombination of virulence factor genes in telomeric chromosome clusters of P. falciparum. Nature 407: 1018Ð1022. doi:10.1038/35039531.

\bibitem{15}	Barry AE, Leliwa-Sytek A, Tavul L, Imrie H, Migot-Nabias F, et al. (2007) Population Genomics of the Immune Evasion (var) Genes of Plasmodium falciparum. PLoS Pathog 3: e34. doi:10.1371/journal.ppat.0030034.st002.

\bibitem{16}	Gupta S, Snow RW, Donnelly CA, Marsh K, Newbold C (1999) Immunity to non-cerebral severe malaria is acquired after one or two infections. Nature medicine 5: 340Ð343.

\bibitem{17}	Bockhorst J, Lu F, Janes JH, Keebler J, Gamain B, et al. (2007) Structural polymorphism and diversifying selection on the pregnancy malaria vaccine candidate VAR2CSA. Molecular and Biochemical Parasitology 155: 103Ð112. doi:10.1016/j.molbiopara.2007.06.007.

\bibitem{18}	Rask TS, Hansen DA, Theander TG, Gorm Pedersen A, Lavstsen T (2010) Plasmodium falciparum Erythrocyte Membrane Protein 1 Diversity in Seven Genomes Ð Divide and Conquer. PLoS Comput Biol 6: e1000933. Available: http://dx.plos.org/10.1371/journal.pcbi.1000933.

\bibitem{19}	Kraemer SM, Smith JD (2006) A family affair: {\it var} genes, PfEMP1 binding, and malaria disease. Current Opinion in Microbiology 9: 374Ð380. doi:10.1016/j.mib.2006.06.006.

\bibitem{20}	Trimnell AR, Kraemer SM, Mukherjee S, Phippard DJ, Janes JH, et al. (2006) Global genetic diversity and evolution of {\it var} genes associated with placental and severe childhood malaria?. Molecular and Biochemical Parasitology 148: 169Ð180. doi:10.1016/j.molbiopara.2006.03.012.

\bibitem{21}	Awadalla P (2003) The evolutionary genomics of pathogen recombination. Nat Rev Genet 4: 50Ð60. doi:10.1038/nrg964.

\bibitem{22}	Apeltsin L, Morris JH, Babbitt, P.C., Ferrin TE (2011) Improving the quality of protein similarity network clustering algorithms using the network edge weight distribution. Bioinformatics 27: 326Ð333.

\bibitem{23}	Bockhorst J, Jojic N (2007) Discovering Patterns in Biological Sequences by Optimal Segmentation. Proceedings of the 23rd International Conference on Uncertainty in Artificial Intelligence.

\bibitem{24}	Newman M (2010) Networks: an introduction.

\bibitem{25}	Alvarez-Ponce D, Lopez P, Bapteste E, McInerney JO (2013) Gene similarity networks provide tools for understanding eukaryote origins and evolution.

\bibitem{26}	Kosakovsky Pond SL (2006) Automated Phylogenetic Detection of Recombination Using a Genetic Algorithm. Molecular Biology and Evolution 23: 1891Ð1901. doi:10.1093/molbev/msl051.

\bibitem{27}	Huson DH, Scornavacca C (2011) A Survey of Combinatorial Methods for Phylogenetic Networks. Genome Biology and Evolution 3: 23Ð35. doi:10.1093/gbe/evq077.

\bibitem{28}	Song YS, Hein J (2005) Constructing minimal ancestral recombination graphs. Journal of Computational Biology 12: 147Ð169.

\bibitem{29}	Halary S, Leigh JW, Cheaib B, Lopez P, Bapteste E (2010) Network analyses structure genetic diversity in independent genetic worlds. Proceedings of the National Academy of Sciences 107: 127Ð132. doi:10.1073/pnas.0908978107.

\bibitem{30}	Fondi M, Fani R (2010) The horizontal flow of the plasmid resistome: clues from inter-generic similarity networks. Environmental Microbiology 12: 3228Ð3242. doi:10.1111/j.1462-2920.2010.02295.x.

\bibitem{31}	Dagan T, Artzy-Randrup Y, Martin W (2008) Modular networks and cumulative impact of lateral transfer in prokaryote genome evolution. Proceedings of the National Academy of Sciences 105: 10039Ð10044.

\bibitem{32}	Bapteste E, Lopez P, Bouchard F (2012) Evolutionary analyses of non-genealogical bonds produced by introgressive descent. doi:10.1073/pnas.1206541109/-/DCSupplemental/pnas.201206541SI.pdf.

\bibitem{33}	Smith JD, Subramanian G, Gamain B, Baruch DI, Miller LH (2000) Classification of adhesive domains in the Plasmodium falciparum erythrocyte membrane protein 1 family. Molecular and Biochemical Parasitology 110: 293Ð310.

\bibitem{34}	Su X-Z, Heatwole VM, Wertheimer SP, Guinet F, Herrfeldt JA, et al. (1995) The large diverse gene family {\it var} encodes proteins involved in cytoadherence and antigenic variation of plasmodium falciparum-infected erythrocytes. Cell 82: 89Ð100.

\bibitem{35}	Gardner MJ, Hall N, Fung E, White O, Berriman M, et al. (2002) Genome sequence of the human malaria parasite Plasmodium falciparum. Nature 419: 498Ð511.

\bibitem{36}	Kraemer SM, Kyes SA, Aggarwal G, Springer AL, Nelson SO, et al. (2007) Patterns of gene recombination shape {\it var} gene repertoires in Plasmodium falciparum: comparisons of geographically diverse isolates. BMC Genomics 8: 45.

\bibitem{37}	Edgar RC (2004) MUSCLE: multiple sequence alignment with high accuracy and high throughput. Nucleic Acids Research 32: 1792Ð1797. doi:10.1093/nar/gkh340.

\bibitem{38}	Castresana J (2000) Selection of conserved blocks from multiple alignments for their use in phylogenetic analysis: 1Ð13.

\bibitem{39}	Karrer B, Newman M (2011) Stochastic blockmodels and community structure in networks. Phys Rev E 83: 016107. doi:10.1103/PhysRevE.83.016107.

\bibitem{40}	Meil? M (2005) Comparing clusterings: an axiomatic view: 577Ð584.

\bibitem{41}	Lavstsen T, Salanti A, Jensen ATR, Arnot DE, Theander TG (2003) Sub-grouping of Plasmodium falciparum 3D7 {\it var} genes based on sequence analysis of coding and non-coding regions. Malar J 2: 27.

\bibitem{42}	Buckee CO, Recker M (2012) Evolution of the Multi-Domain Structures of Virulence Genes in the Human Malaria Parasite, Plasmodium falciparum. PLoS Comput Biol 8: e1002451. doi:10.1371/journal.pcbi.1002451.g006.

\bibitem{43}	Kraemer SM, Smith JD (2003) Evidence for the importance of genetic structuring to the structural and functional specialization of the Plasmodium falciparum {\it var} gene family. Molecular Microbiology 50: 1527Ð1538. doi:10.1046/j.1365-2958.2003.03814.x.

\bibitem{44}	Bopp SER, Manary MJ, Bright AT, Johnston GL, Dharia NV, et al. (2013) Mitotic Evolution of Plasmodium falciparum Shows a Stable Core Genome but Recombination in Antigen Families. PLoS Genet 9: e1003293. doi:10.1371/journal.pgen.1003293.s008.

\bibitem{45}	Newman M (2002) Assortative Mixing in Networks. Phys Rev Lett 89: 208701. doi:10.1103/PhysRevLett.89.208701.

\bibitem{46}	Warimwe GM, Keane TM, Fegan G, Musyoki JN, Newton CRJC, et al. (2009) Plasmodium falciparum {\it var} gene expression is modified by host immunity. Proceedings of the National Academy of Sciences 106: 21801Ð21806. Available: http://www.pnas.org/content/106/51/21801.full.pdf+html.

\bibitem{47}	Bull PC, Kyes S, Buckee CO, Montgomery J, Kortok MM, et al. (2007) An approach to classifying sequence tags sampled from Plasmodium falciparum {\it var} genes. Molecular and Biochemical Parasitology 154: 98Ð102. doi:10.1016/j.molbiopara.2007.03.011.

\bibitem{48}	Warimwe GM, Fegan G, Musyoki JN, Newton CRJC, Opiyo M, et al. (2012) Prognostic Indicators of Life-Threatening Malaria Are Associated with Distinct Parasite Variant Antigen Profiles. Science Translational Medicine 4: 129ra45Ð129ra45. Available: http://www.translationalmedicine.org/content/4/129/129ra45.short.

\bibitem{49}	ORGEL LE (1963) The maintenance of the accuracy of protein synthesis and its relevance to ageing. Proceedings of the National Academy of Sciences 49: 517Ð521.

\bibitem{50}	Mucha PJ, Richardson T, Macon K, Porter MA, Onnela JP (2010) Community Structure in Time-Dependent, Multiscale, and Multiplex Networks. Science 328: 876Ð878. doi:10.1126/science.1184819.

\end{thebibliography}

\begin{thebibliography}{99}

\bibitem{Edgar:2004bo} Edgar RC (2004) MUSCLE: multiple sequence alignment with high accuracy and high throughput. Nucleic Acids Research 32: 1792–1797. doi:10.1093/nar/gkh340.

\bibitem{Darriba:2011ht} Darriba D, Taboada GL, Doallo R, Posada D (2011) ProtTest 3: fast selection of best-fit models of protein evolution. Bioinformatics 27: 1164–1165. doi:10.1093/bioinformatics/btr088.

\bibitem{Stamatakis:2006de} Stamatakis A (2006) RAxML-VI-HPC: maximum likelihood-based phylogenetic analyses with thousands of taxa and mixed models. Bioinformatics 22: 2688–2690. doi:10.1093/
bioinformatics/btl446.

\bibitem{Huson:1998um} Huson DH (1998) SplitsTree: analyzing and visualizing evolutionary data. Bioinformatics 14: 68–73.
\end{thebibliography}
\end{document}